\documentclass[12pt]{article}
\usepackage{graphicx}
\usepackage{amssymb}
\usepackage{amsmath}
\allowdisplaybreaks

\newlength{\extraspace}
\newlength{\extraspaces}
\newcommand{\be}{\begin{equation}
\addtolength{\abovedisplayskip}{\extraspaces}
\addtolength{\belowdisplayskip}{\extraspaces}
\addtolength{\abovedisplayshortskip}{\extraspace}
\addtolength{\belowdisplayshortskip}{\extraspace}}
\newcommand{\ee}{\end{equation}}

\newcommand{\ba}{\begin{eqnarray}
\addtolength{\abovedisplayskip}{\extraspaces}
\addtolength{\belowdisplayskip}{\extraspaces}
\addtolength{\abovedisplayshortskip}{\extraspace}
\addtolength{\belowdisplayshortskip}{\extraspace}}
\newcommand{\ea}{\end{eqnarray}}

\newcommand{\bd}{\begin{displaymath}
\addtolength{\abovedisplayskip}{\extraspaces}
\addtolength{\belowdisplayskip}{\extraspaces}
\addtolength{\abovedisplayshortskip}{\extraspace}
\addtolength{\belowdisplayshortskip}{\extraspace}}
\newcommand{\ed}{\end{displaymath}}

\newcommand{\newsection}[1]{
\vspace{12mm}
\pagebreak[3]
\addtocounter{section}{1}
\setcounter{equation}{0}
\setcounter{subsection}{0}
\noindent{\bf \thesection. #1}
\nopagebreak
\medskip
\nopagebreak}

\newcommand{\newsubsection}[1]{
\vspace{0.8cm}
\pagebreak[3]
\addtocounter{subsection}{1}
\noindent{\it \thesubsection. #1}
\nopagebreak
\vspace{2mm}
\nopagebreak}

\newcounter{saveeqn}

\newcommand{\dif}{\mathrm{d}}
\newcommand{\me}{\mathrm{e}}

\begin{document}
\addtolength{\baselineskip}{1.5mm}

\thispagestyle{empty}
\begin{flushright}

\end{flushright}
\vbox{}
\vspace{2cm}

\begin{center}
{\LARGE{A doubly rotating black ring with dipole charge
        }}\\[16mm]
{Yu Chen,$^1$~~Kenneth Hong$^1$~~and~~Edward Teo$^{1,2}$}
\\[6mm]
$^1${\it Department of Physics,
National University of Singapore, 
Singapore 119260}\\[5mm]
$^2${\it Centre for Gravitational Physics, College of Physical and Mathematical Sciences,\\[1mm]
The Australian National University, Canberra ACT 0200, Australia}\\[15mm]

\end{center}
\vspace{2cm}

\centerline{\bf Abstract}
\bigskip
\noindent
We present a dipole-charged generalisation of the Pomeransky--Sen'kov black ring in five-dimensional Kaluza--Klein theory. It rotates in two independent directions, although one of the rotations has been tuned to achieve balance, so that the space-time does not contain any conical singularities. This solution was constructed using the inverse-scattering method in six-dimensional vacuum gravity. We then study various physical properties of this solution, with particular emphasis on the new features that the dipole charge introduces.


\newpage

\newsection{Introduction}

The 2001 discovery of a five-dimensional rotating black ring by Emparan and Reall \cite{Emparan:2001wn} heralded in an exciting era for the study of higher-dimensional black holes. Its $S^1\times S^2$ horizon topology distinguishes it from the more usual spherical one of the Myers--Perry black hole \cite{Myers:1986un}, and is responsible for a number of interesting features of this solution. For example, the rotation in the $S^1$ direction creates a centrifugal force that is able to balance the self-gravity of the ring, giving rise to a space-time without conical singularities. It is also possible for a black ring to rotate in the azimuthal direction of the $S^2$. Such an $S^2$-rotating black ring, without any rotation in the $S^1$ direction, was discovered in \cite{Mishima:2005id,Figueras:2005zp}. The balanced doubly rotating black ring, with rotations in both the $S^1$ and $S^2$ directions, was discovered by Pomeransky and Sen'kov \cite{Pomeransky:2006bd}. The most general doubly rotating black ring, in which the balance condition is not enforced, was found in \cite{Morisawa:2007di,Chen:2011jb}.

It is of obvious interest to generalise these vacuum black ring solutions to include charge. This would allow the embedding and study of black rings in string theory, among other possibilities. Like five-dimensional black holes, black rings can carry a conserved electric charge with respect to a two-form field strength. For example, if we consider the following generic Einstein--Maxwell-dilaton action:
\be\label{5D_action}
S=\frac{1}{16\pi G_5}\int\dif^5x\sqrt{-g}\,\big(R-\hbox{$\frac{1}{2}$}\partial_\mu\varphi\partial^\mu\varphi-\hbox{$\frac{1}{4}$}\me^{-\alpha\varphi}F_{\mu\nu}F^{\mu\nu}\big)\,,
\ee
then the electric charge of the system is defined by
\be
Q=\frac{1}{4\pi}\int_{S^3} \me ^{-\alpha\varphi}\star F\,,
\ee
where the integration is taken over the 3-sphere of a constant time slice at infinity. Happily, the standard charging transformations developed for black holes can also be applied to black rings. 
Charged black rings in various theories, including string theory, have been considered in, say \cite{Elvang:2003yy,Elvang:2003mj,Bouchareb:2007ax,Hoskisson:2008qq,Gal'tsov:2009da}.

Unlike black holes however, black rings can carry a new type of magnetic charge by virtue of their horizon topology. It is locally defined by
\be
\label{definition_dipole_charge}
\mathcal{Q}=\frac{1}{4\pi}\int _{S^2} F\,,
\ee
where the integration is taken over a 2-sphere which encloses the $S^2$-section of the black ring horizon. The sign of $\mathcal{Q}$ depends on a choice of orientation of this 2-sphere. It follows that $\mathcal{Q}$ has opposite signs for 2-spheres on opposite sides of the ring horizon, since the orientation induced on these two 2-spheres by the ring horizon are opposite to each other \cite{Dowker:1995sg}.
For this reason, it is known as a ``dipole charge'', although it does not obey any conservation law. The first example of a dipole-charged black ring was discovered by Emparan \cite{Emparan:2004wy}, as a solution to the Einstein--Maxwell-dilaton action (\ref{5D_action}) with arbitrary dilaton coupling $\alpha$. As it rotates in the $S^1$ direction, it generalises the Emparan--Reall black ring.

It is natural to wonder if a dipole black ring including rotation in the $S^2$ direction can be found. This has, unfortunately, proven to be a difficult problem, as the known solution-generating techniques, such as those developed in \cite{Yazadjiev:2006hw,Yazadjiev:2006ew}, do not seem to apply to this case. Even the inverse-scattering method (ISM) \cite{Belinski:2001,Pomeransky:2005sj}, which has been very successfully used to generate vacuum black ring solutions \cite{Emparan:2008eg,Iguchi:2011qi}, is not directly applicable to the charged case. There have been attempts to generalise the ISM to include charge (e.g., \cite{Figueras:2009mc}). A recent breakthrough, however, came with the work of Rocha et al.~\cite{Rocha:2011vv}, who realised that the ISM is applicable to the Einstein--Maxwell-dilaton action (\ref{5D_action}) when $\alpha=2\sqrt{2/3}$. This special case corresponds to a five-dimensional Kaluza--Klein theory, in which (\ref{5D_action}) can be obtained by a circle reduction of six-dimensional vacuum gravity using the ansatz
\be
\label{ansatz}
\dif s_6^2=\me^{\frac{\varphi}{\sqrt{6}}}\,\dif s_5^2+\me^{-\sqrt{\frac{3}{2}}\,\varphi}(\dif w+A)^2,
\ee
where $w$ is the coordinate to be reduced along.
The ISM can of course be applied in six dimensions in the present case, and Rocha et al.\ showed how it can be used to generate Emparan's dipole black ring starting from a suitable seed solution.

It was suggested in \cite{Rocha:2011vv} that this method can also be used to generate a dipole black ring including $S^2$ rotation, although it was left as an open problem. The purpose of this paper is to address this issue. Indeed, we will show that the ISM can be applied on a suitable seed solution in six dimensions, to generate a doubly rotating dipole black ring of five-dimensional Kaluza--Klein theory. For simplicity, we concentrate only on the balanced case, so our solution may be considered a dipole-charged generalisation of the Pomeransky--Sen'kov black ring. 

We then study the physical properties of this black ring, emphasizing the new features that the dipole charge introduces. In particular, we show that our solution satisfies a version of the first law of black-hole thermodynamics that includes dipole charge, which was proved by Copsey and Horowitz \cite{Copsey:2005se} (see also \cite{Astefanesei:2005ad}). We also describe its effect on the phase space of the black ring. Finally, we show in detail how various limits of this solution can be taken.

This paper is organised as follows: In Sec.~2, the ISM construction of the doubly rotating dipole black ring solution is described. The solution is explicitly written down in Sec.~3, and its physical properties analysed in Sec.~4. The phase space structure is discussed in Sec.~5, while the various limits of this solution are presented in Sec.~6. The paper ends with a brief discussion in Sec.~7.

\newsection{ISM construction}

In \cite{Rocha:2011vv}, Rocha et al.\ showed how the inverse-scattering method can be used to generate the singly rotating dipole black ring in five-dimensional Kaluza--Klein theory. They did this by applying a two-soliton transformation on a certain six-dimensional seed solution. One of the soliton transformations is responsible for introducing a five-dimensional dipole charge, while the other is responsible for introducing $S^1$-rotation to the ring.

Separately, it is known how the ISM can be used to obtain the doubly rotating vacuum black ring \cite{Pomeransky:2006bd}. This requires a three-soliton transformation on a certain (five-dimensional) seed solution. One of them is responsible for introducing the $S^1$-rotation, while the other two are needed to introduce the $S^2$-rotation. This procedure was explained in detail in \cite{Chen:2011jb}.

It is therefore relatively straightforward to combine the ISM procedures of \cite{Rocha:2011vv} and \cite{Chen:2011jb} into a single procedure to generate a doubly rotating dipole black ring. This requires a four-soliton transformation on the same six-dimensional seed as used in \cite{Rocha:2011vv}: one soliton to introduce dipole charge, and the other three to make the black ring rotate in two directions. In the rest of this section, we will describe the essential points of this construction. To keep the technical details to a minimum, some familiarity with the ISM will be assumed of the reader. Reviews of the ISM relevant to five-dimensional black holes and black rings may be found in, say \cite{Pomeransky:2005sj,Emparan:2008eg,Iguchi:2011qi}.

\begin{figure}[t]
\begin{center}
\includegraphics{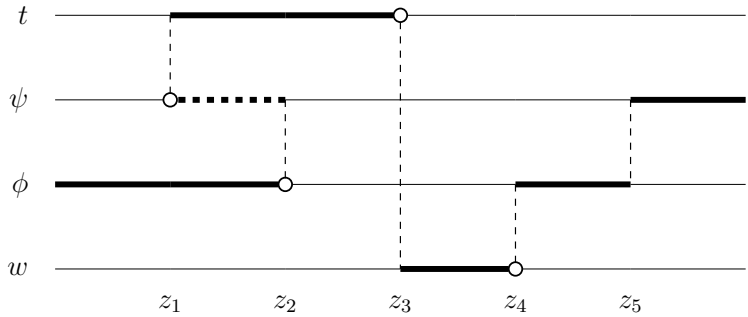}
\caption{The rod sources of the seed solution for the doubly rotating dipole black ring when lifted to six dimensions. The thin lines denote the $z$-axis and the thick lines denote rod sources of mass $\frac{1}{2}$ per unit length along this axis. The thick dashed line denotes a rod source with negative mass density $-\frac{1}{2}$. Small circles represent the operations of removing solitons from the seed, each with a BZ vector having a single non-vanishing component along the coordinate that labels the $z$-axis where the circle is placed. \label{rod_seed}}
\label{seed_rod}
\end{center}
\end{figure}

We begin with a  seed solution having the rod structure as shown in Fig.~\ref{seed_rod}.\footnote{We remark that one can also start with the following seed solution:
\begin{equation*}
G_0={\textrm{diag}}\, \bigg\{ -{\frac {{\mu_1}}{{\mu_3}}},\frac{\mu_2\mu_5}{\mu_1}, \frac{\rho^2\mu_3}{\mu_2\mu_4}, \frac{\mu_4}{\mu_5}\bigg\}\,,
\end{equation*}
in which the two finite space-like rods with positive mass density are swapped, and generate the same final solution up to coordinate transformations and parameter redefinitions.} In Weyl--Papapetrou coordinates \cite{Emparan:2001wk}
\be
\dif s_6^2=G_{ab}\,\dif x^a\dif x^b+\me^{2\gamma}(\dif\rho^2+\dif z^2)\,,
\ee
its $G$-matrix and corresponding conformal factor are given by
\ba
G_0&=&{\mathrm{diag}}\,\bigg\{-{\frac{{\mu_1}}{{\mu_3}}}\,,\frac{\mu_2\mu_5}{\mu_1}\,,\frac{\rho^2\mu_4}{\mu_2\mu_5}\,,\frac{\mu_3}{\mu_4}\bigg\}\,,\cr
\label{gamma0}
\me^{2\gamma_0}&=&k^2\,\frac{\mu_2\mu_5R_{12}R_{13}R_{15}R_{24}R_{34}R_{45}}{\mu_1R_{25}^2R_{11}R_{22}R_{33}R_{44}R_{55}}\,.
\ea
Here, $\mu_i\equiv\sqrt{\rho^2+(z-z_i)^2}-(z-z_i)$, $R_{ij}\equiv\rho^2+\mu_i\mu_j$, and $k$ is an arbitrary integration constant. 
Using the ISM, we then perform the following four-soliton transformation on this seed:
\begin{enumerate}
\item Remove a soliton at each of $z_1$, $z_2$, $z_3$ and $z_4$, with trivial Belinski--Zakharov (BZ) vectors $(0,1,0,0)$, $(0,0,1,0)$, $(1,0,0,0)$ and $(0,0,0,1)$, respectively;
\item Add back a soliton at each of $z_1$, $z_2$, $z_3$ and $z_4$, with non-trivial  BZ vectors $(C_1,1,0,0)$, $(0,C_2,1,0)$, $(1,0,0,C_3)$ and $(0,0,C_4,1)$, respectively. Here, $C_1$, $C_2$, $C_3$ and $C_4$ are the new, so-called BZ parameters.
\end{enumerate}
We note that if $C_2=C_3=0$, we recover the ISM procedure of \cite{Rocha:2011vv}. On the other hand, if $C_4=0$ and $z_3=z_4$, we effectively recover the ISM procedure of \cite{Chen:2011jb}.

In the first step above, the act of removing a soliton at $z=z_k$ with a trivial BZ vector having a non-vanishing $a$-th component, refers to multiplying the diagonal element $(G_{0})_{aa}$ of the seed solution by a factor $-\frac{\mu_k^2}{\rho^2}$. So in the current case, after the first step, we obtain the new $G$-matrix:
\be
\tilde{G}_0= {\textrm{diag}}\,  \bigg\{\frac{\mu_1 \mu_3}{\rho^2} ,-{\frac {{\mu_1}{\mu_2}\mu_5}{{\rho}
^{2}}},-{\frac {{\mu_2}{\mu_4}}{{\mu_5}}},-{\frac {{\mu_3}{\mu_4}}{{\rho^2}}} \bigg\}\,.
\ee
The generating matrix $\tilde{\Psi}_0$ can be obtained directly by performing the following replacements to $\tilde{G}_0$: $\mu_i\rightarrow \mu_i-\lambda$ and $\rho^2\rightarrow \rho^2-2z\lambda-\lambda^2$, where $\lambda$ is a spectral parameter. One can then easily follow \cite{Pomeransky:2005sj} to carry out the second step above. In computing the vectors $m^{(k)}$, we used the same trick as was described in \cite{Chen:2011jb}. The conformal factor can also be easily calculated:
\be
\label{final_f}
\me^{2\gamma}=\me^{2\gamma_0}\frac{\det\Gamma(C_1,C_2,C_3,C_4)}{\det\Gamma(C_1=0,C_2=0,C_3=0,C_4=0)}\,,
\ee
where the matrix $\Gamma$, as defined in \cite{Pomeransky:2005sj}, is that corresponding to the second step above.

Once the new solution has been generated, we can calculate its rod structure, following the prescription of \cite{Chen:2010zu,Chen:2010ih} (see also \cite{Harmark:2004rm,Hollands:2007aj}). Counting the rods from the left, we then join up Rods 1 and 2, as well as Rods 4 and 5, by requiring that Rod 1 has the same (normalised) direction as Rod 2, and Rod 4 has the same direction as Rod 5. These conditions give equations for $C_1{}^2$ and $C_4{}^2$. Without loss of generality, we take the solution
\be
\label{C1andC4}
C_1=\sqrt{\frac{z_{31}}{2z_{21}z_{51}}}\,,\qquad C_4=\sqrt{\frac{2z_4^2z_{42}}{z_{43}z_{54}}}\,,
\ee
where $z_{ij}\equiv z_i-z_j$. By setting these values of the BZ parameters, we effectively eliminate the turning points $z_1$ and $z_4$; this will leave the resulting solution with three genuine turning points. 

At this stage, we transform from Weyl--Papapetrou coordinates $(\rho,z)$ to C-metric-like coordinates $(x,y)$ \cite{Harmark:2004rm}:
\be
\rho^2=\frac{4\varkappa^4(1-x^2)(y^2-1)(1+cx)(1+cy)}{(x-y)^4}\,,\qquad
z=\frac{\varkappa^2(1-xy)(2+cx+cy)}{(x-y)^2}\,,
\ee
with the locations of the five turning points fixed to be
\be
z_1=-d\varkappa^2,\qquad z_2=-c\varkappa^2,\qquad
z_3=c\varkappa^2,\qquad z_4=e\varkappa^2,\qquad z_5=\varkappa^2.
\ee
Since the solution after imposing (\ref{C1andC4}) has just three genuine turning points, its metric can be written in terms of purely algebraic expressions of $x$ and $y$. Some technical details can be found in \cite{Chen:2011jb} on how to convert the solution from Weyl--Papapetrou coordinates to C-metric-like coordinates. This step simplifies the solution dramatically.

Now, to ensure that the solution describes a balanced doubly rotating dipole black ring in five dimensions, we also require that after dimensional reduction along $w$, Rods 1 and 4 have the same direction. The conditions for this to be true are
\be
\label{z1andC}
C_2=-\sqrt{\frac{(d+c)(e+c)}{(d-c)(e-c)(1+d)(1-e)^3}}\,\frac{(1+c)^3}{c}\,C_3\,,\qquad d=\frac{c^2+2c+e}{1-e}\,.
\ee
We note that the second condition could have been obtained by setting $C_2=C_3=0$. Essentially, it means that the balanced doubly rotating dipole black ring can also be directly generated from the balanced singly rotating dipole black ring; a similar situation has been observed in the construction of the Pomeransky--Sen'kov black ring \cite{Pomeransky:2006bd}.

The next step involves rotating the solution to standard orientation \cite{Chen:2010zu}, where the first and last space-like rods have directions $(0,0,1,0)$ and $(0,1,0,0)$, respectively. This will ensure that the metric takes a simple diagonal form at infinity, and is accomplished by a linear transformation of the $G$-matrix: $G'=A^TGA$, with a suitable choice of matrix $A$ obeying $|\det A|=1$.

Finally, we have found that the metric components take the simplest form when $C_3$ and $e$ are eliminated in favour of the new parameters $b$ and $a$ as follows:
\be
C_3=\sqrt{\frac{b(1-e)(e-c)}{3c-ce+e+c^2}}\,,
\qquad e=\frac{2a+ac-c}{1+a}\,.
\ee
After making a suitable choice of the integration constant $k$ in (\ref{gamma0}) to ensure asymptotic flatness, the metric of the resulting solution is given below in Eq.~(\ref{solution6D}) in terms of the final physical parameters $a$, $b$, $c$, and $\varkappa$.

\newsection{The solution and rod structure}

The balanced doubly rotating dipole black ring solution lifted to six dimensions can be expressed in the following form:
\ba
\label{solution6D}
\dif s^2_6&=&\frac{K(x,y)}{H(x,y)}\left(\dif w+A_{t}\,\dif t+A_{\psi}\,\dif\psi+A_{\phi}\,\dif\phi\right)^2-\frac{H(y,x)}{K(x,y)}\left(\dif t+\omega_1\,\dif\psi+\omega_2\,\dif\phi\right)^2\cr
&&+\frac{2\varkappa^2H(x,y)}{(x-y)^2}\,\bigg\{\frac{F(x,y)\left(\dif\psi+\omega_3\,\dif\phi\right)^2}{H(x,y)H(y,x)}-\frac{G(x)G(y)\,\dif\phi^2}{F(x,y)}+\frac{1}{\Phi\Psi}\bigg[\frac{\dif x^2}{G(x)}-\frac{\dif y^2}{G(y)}\bigg]\bigg\}\,,~~~~~~
\ea
where
\ba
\label{functionsA}
A_{t}&=&-\sqrt{b(a^2-c^2)(1-a^2)}\,\frac{c(1+b)(1-xy)(x-y)}{K(x,y)}\,,\cr
A_{\psi}&=&-\sqrt{\frac{2ab(a-c)(1-a^2)}{\Phi\Psi}}\,\frac{\varkappa c(1+b)(1+y)}{K(x,y)}\,\cr
&&\times\left[x(1-y)(1+c)\Phi+(1-x)^2(a+ab+bcy+c)\right],\cr
A_{\phi}&=&-\sqrt{\frac{2a(a-c)}{\Phi\Psi}}\,\frac{\varkappa(1+b)(1+x)L(x,y)}{K(x,y)}\,,\cr
\label{functionsOmega}
\omega_1&=&\sqrt{\frac{2a(a+c)}{\Phi\Psi}}\,\frac{\varkappa(1+b)(1+y)J_{+}(x,y)}{H(y,x)}\,,\cr
\omega_2&=&\sqrt{\frac{2ab(a+c)(1-a^2)}{\Phi\Psi}}\,\frac{\varkappa c(1+b)(1-x^2)}{H(y,x)}\,\left[(1+cy)(a+ab+by)-c-y\right],\cr
\omega_3&=&\frac{\sqrt{b(1-a^2)}}{\Phi\Psi}\,\frac{ac(1+b)(x-y)(1-x^2)(1-y^2)}{F(x,y)}\cr
&&\times\left[b(1+cx)(1+cy)(1-b-a^2-a^2b)-(1-c^2)(1-b+a^2+a^2b)\right],\qquad
\ea
and the functions $G$, $K$, $H$, $F$, $L$ and $J_\pm$ ($J_{-}$ is defined for use below) are given by
\ba
\label{functions1}
G(x)&=&(1-x^2)(1+cx)\,,\cr
K(x,y)&=&-a^2(1+b)\left[bx^2(1+cy)^2+(c+x)^2\right]+\left[b(1+cy)-1-cx\right]^2+bc^2(1-xy)^2,\cr
H(x,y)&=&-a^2(1+b)\left[b(1+cx)(1+cy)xy+(c+x)(c+y)\right]-a(1+b)(x-y)\big[c^2-1\cr
&&+b(1+cx)(1+cy)\big]+\left[b(1+cy)-1-cx\right]\left[b(1+cx)-1-cy\right]+bc^2(1-xy)^2,\cr
F(x,y)&=&\frac{1-y^2}{\Phi\Psi}\,\bigg\{bcG(x)\Big\{c(y^2-1)\left[a^2(1+b)-b+1\right]^2-4a^2y(1-b^2)(1+cy)\Big\}\cr
&&-(1+cy)\Big\{a^2(1+b)^2\left[a^2(c+x+bx+bcx^2)^2-(c+x-bx-bcx^2)^2\right]\cr
&&-(1-b)^2(1+cx)^2\left[a^2(1+b)^2-(1-b)^2\right]\Big\}\bigg\}\,,\cr
L(x,y)&=&a^2(1+b)\left[bx(1+cy)^2+(1+c)(c+x)\right]-a(1-x)\left[b^2(1+cy)^2+c^2-1\right]\cr
&&-\left[b(1+cy)-c-1\right]\left[b(1+cy)-cx-1\right]-bc^2(1-y)(1-xy)\,,\cr
J_{\pm}(x,y)&=&a^2(1+b)\left[bx(1+cx)(1+cy)+(1+c)(c+x)\right]\cr
&&\pm a\left\{(1-x)\left[b(1+cx)+c-1\right]\left[b(1+cy)+c+1\right]-2bc(1-y)(1+cx)\right\}\cr
&&-\left[b(1+cx)-c-1\right]\left[b(1+cy)-cx-1\right]-bc^2(1-x)(1-xy)\,.
\ea
To simplify the above expressions, we have introduced the following abbreviations:
\be
\Phi\equiv1+a-b+ab\,,\qquad\Psi\equiv1-a-b-ab\,.
\ee

The coordinates take the ranges $-\infty<t<\infty$, $0\le\psi,\phi<2\pi$ and $-\infty<y\le-1\le x\le 1$. There are four independent parameters, $a$, $b$, $c$ and $\varkappa$, satisfying the constraints
\be
\label{param_ranges}
0\le c\le a<1\,,\qquad0\le b<\frac{1-a}{1+a}\,,\qquad\varkappa>0\,.
\ee
It can be checked that the former two constraints ensure that the quantities $\Phi$ and $\Psi$ are positive. As will be clear below, the parameters have the following interpretations: roughly speaking, $\varkappa$ sets the scale of the solution, $c$ characterises the size of the black hole, $b$ controls the $S^2$ rotation, while $a$ controls the dipole charge.

It is instructive to calculate the rod structure of the above solution in its six-dimensional form, following \cite{Chen:2010zu,Chen:2010ih}. 
It possesses four mutually commuting Killing vector fields $\big(\frac{\partial}{\partial t},\frac{\partial}{\partial \psi},\frac{\partial}{\partial \phi},
\break\frac{\partial}{\partial w}\big)$,
which will be used as a basis to express the rod directions.
There are three turning points in the rod structure, located at $(\rho=0,z=z_1\equiv-c\varkappa^2)$ or $(x=-1,y=-\frac{1}{c})$, $(\rho=0,z=z_2\equiv c\varkappa^2)$ or $(x=1,y=-\frac{1}{c})$, and $(\rho=0,z=z_3\equiv\varkappa^2)$ or $(x=1,y=-1)$, respectively.
They divide the $z$-axis into four rods:
\begin{itemize}
\item Rod 1:\enskip a semi-infinite space-like rod at $(\rho=0, z\leq z_1)$ or $(x=-1, -\frac{1}{c}\le y<-1)$, with direction $\ell_1=(0,0,1,0)$.
\item Rod 2:\enskip a finite time-like rod at $(\rho=0, z_1\leq z\leq z_2)$ or $(-1\le x\le 1, y=-\frac{1}{c})$, with direction $\ell_2=\frac{1}{\kappa}(1,\Omega_\psi,\Omega_\phi,\Omega_w)$, where
\ba
\label{kappa}
\kappa&=&\frac{1}{4\varkappa(1+b)}\,\sqrt{\frac{2\Phi\Psi^3}{a(a+c)(1-a^2)}}\,,\cr
\label{Omega_psi}
\Omega_\psi&=&\frac{1}{\varkappa}\,\sqrt{\frac{a\Psi}{2(a+c)\Phi}}\,,\cr
\label{Omega_phi}
\Omega_\phi&=&\frac{1-b+a^2+a^2b}{\varkappa(1+b)}\,\sqrt{\frac{b\Psi}{2a(a+c)(1-a^2)\Phi}}\,,\cr
\Omega_w &=&-\frac{\Psi}{\Phi}\sqrt{\frac{b(1-a)(a-c)}{(1+a)(a+c)}}\,.
\ea
\item Rod 3:\enskip a finite space-like rod at $(\rho=0, z_2\leq z\leq z_3)$ or $(x=1,-\frac{1}{c}\le y\le-1)$, with direction
\be
\ell_3=\bigg(0,0,1,-\frac{2\varkappa(1+b)\sqrt{2a(a-c)}}{\sqrt{\Phi\Psi}}\bigg).
\ee
\item Rod 4:\enskip a semi-infinite space-like rod at $(\rho=0, z\geq z_3)$ or $(-1<x\le1, y=-1)$, with direction $\ell_4=(0,1,0,0)$.
\end{itemize}
Note in particular that Rod 3 has a non-zero component in the $w$-direction. It is what will eventually give rise to a non-zero magnetic dipole charge when the solution is reduced to five dimensions, as first observed in \cite{Rocha:2011vv}. But now, Rod 2 also has a non-zero component in the $w$-direction. This component is a manifestation of the electric quadrupole moment of the reduced solution. 

Now performing dimensional reduction on (\ref{solution6D}) using the ansatz (\ref{ansatz}), we obtain a solution of five-dimensional Kaluza--Klein theory. The resulting metric takes the following form:
\ba
\label{solution5D}
\dif s^2_5&=&-\left[\frac{H(y,x)^3}{K(x,y)^2H(x,y)}\right]^{\frac{1}{3}}\left(\dif t+\omega_1\,\dif\psi+\omega_2\,\dif\phi\right)^2+\frac{2\varkappa^2}{(x-y)^2}\left[K(x,y)H(x,y)^2\right]^{\frac{1}{3}}\cr
&&\times\bigg\{\frac{F(x,y)\left(\dif\psi+\omega_3\,\dif\phi\right)^2}{H(x,y)H(y,x)}-\frac{G(x)G(y)\,\dif\phi^2}{F(x,y)}+\frac{1}{\Phi\Psi}\bigg[\frac{\dif x^2}{G(x)}-\frac{\dif y^2}{G(y)}\bigg]\bigg\}\,,\quad
\ea
where the functions $G$, $K$, $H$ and $F$ are given in (\ref{functions1}), and $\omega_{1,2,3}$ are given in (\ref{functionsOmega}). The gauge potential $A$ is
\be
\label{A_soln}
A=A_{t}\,\dif t+A_{\psi}\,\dif\psi+A_{\phi}\,\dif\phi\,,
\ee
where $A_{t,\psi,\phi}$ are defined as in (\ref{functionsA}) and dilaton field $\varphi$ is given by
\be
\label{phi_soln}
\me^{-\varphi}=\left[\frac{K(x,y)}{H(x,y)}\right]^{\sqrt{\frac{2}{3}}}.
\ee

For completeness, we also present the five-dimensional solution when the two-form field strength $F$ is dualised to a three-form field strength $H$ via
\be
\label{duality}
H={\rm e}^{-\alpha\varphi}\star F\,,\qquad\tilde\varphi=-\varphi\,.
\ee
 The corresponding two-form potential $B$ is given by
\be
B=B_{t\psi}\,\dif t\wedge\dif\psi+B_{t\phi}\,\dif t\wedge\dif\phi+B_{\phi\psi}\,\dif\phi\wedge\dif\psi\,,
\ee
where
\ba
\label{B_field_components}
B_{t\psi}&=&-\sqrt{\frac{2a(a-c)}{\Phi\Psi}}\frac{\varkappa (1+b)(1+y)J_{-}(x,y)}{H(x,y)}\,,\cr
B_{t\phi}&=&-\sqrt\frac{2ab(1-a^2)(a-c)}{\Phi\Psi}\frac{c\varkappa(1+b)(1-x^2)[c+y+(1+cy)(a+ab-by)]}{H(x,y)}\,,\cr
B_{\phi\psi}&=&\frac{2c\varkappa^2(1+b)\sqrt{b(1-a^2)(a^2-c^2)}}{\Psi(x-y)H(x,y)}\,(1-x^2)(1+y)[a(1+b)(1-x)(1+cy)\cr
&&+(1-y)(b+bcy-1-cx)]\,.
\ea
Note that the integration constants of the above three components of $B$ are chosen such that $B_{t\psi}$ and $B_{\phi\psi}$ vanish on the axis $y=-1$, while $B_{t\phi}$ vanishes on the axis $x=-1$. It follows that $B$ vanishes at infinity, since the axes $x=-1$ and $y=-1$ extend to infinity, and $B$ is constant there \cite{Copsey:2005se}.

\begin{figure}[t]
\begin{center}
\includegraphics{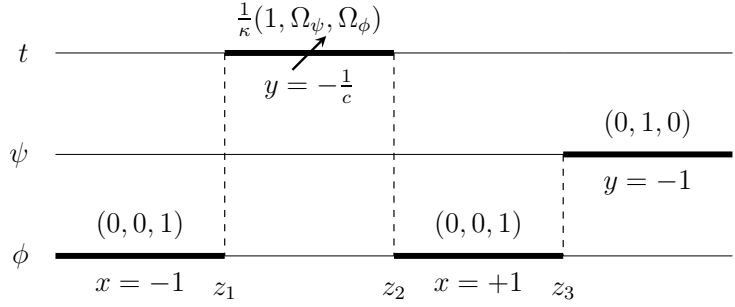}
\caption{The rod structure of the balanced doubly rotating dipole black ring. The location of each rod is indicated below it, while the direction is indicated above it. The arrow on the horizon rod indicates that its direction vector has components in the $\psi$ and $\phi$ directions as well.}
\label{Final rod structure}
\end{center}
\end{figure}

The rod structure of the five-dimensional black ring metric (\ref{solution5D}) can also be calculated. We remark that it is the same as that of (\ref{solution6D}), but now with the $w$-components of all the rod directions removed. This rod structure is shown schematically in Fig.~\ref{Final rod structure}, and it clearly describes a balanced doubly rotating (dipole) black ring. In the following section, we shall examine the physical properties of this black ring in more detail.

\newsection{Physical properties}

We begin by noting that the metric of the doubly rotating dipole black ring (\ref{solution5D}) is asymptotically flat, with infinity located at $(x,y)\rightarrow(-1,-1)$. This can be explicitly seen by changing coordinates $\{x=-1+\frac{4\varkappa^2}{r^2}(1-c)\cos^2\theta,y=-1-\frac{4\varkappa^2}{r^2}(1-c)\sin^2\theta\}$.
The ADM mass $M$ and angular momenta $J_\psi$, $J_\phi$ of the space-time can then be calculated to be
\ba
M&=&\frac{\pi\varkappa^2(1+b)[(a+c)\Phi+a(1-b+c+bc)]}{G_5\Phi\Psi}\,,\cr
J_\psi&=&\frac{2\pi\varkappa^3(1+b)[(1+c)\Phi+2bc(1-a)]}{G_5\Psi^{3/2}}\,\sqrt{\frac{a(a+c)}{2\Phi}}\,,\cr
J_\phi&=&\frac{2\pi\varkappa^3c(1+b)}{G_5\Psi^{3/2}}\,\sqrt{\frac{2ab(a+c)(1-a^2)}{\Phi}}\,.
\ea

As can be seen from the rod structure, there is an event horizon located at $y=-\frac{1}{c}$. It has a ring topology $S^1\times S^2$, with $\frac{\partial}{\partial\psi}$ generating the $S^1$ and $\frac{\partial}{\partial\phi}$ generating the rotational symmetry of the $S^2$. The surface gravity $\kappa$, and angular velocities $\Omega_\psi$ and $\Omega_\phi$, of the event horizon are given in (\ref{kappa}). Its area is
\be
A_{\rm H}=16\pi^2\varkappa^3c(1+b)\sqrt{\frac{2a(a+c)(1-a^2)}{\Phi\Psi^3}}\,.
\ee

It is straightforward to check that this system does not carry conserved electric charge, i.e., $Q=0$. It does, however, carry a magnetic dipole charge, which can be calculated to be
\be
\mathcal{Q}=\frac{\varkappa(1+b)\sqrt{2a(a-c)}}{\sqrt{\Psi\Phi}}\,.
\ee
Note that this charge can be expressed in terms of the rod structure of the (six-dimensional) solution (\ref{solution6D}) as follows:
\be
\label{dipole_formula}
\mathcal{Q}=\hbox{$\frac{1}{2}$}\left(\ell_1 [w]-\ell_3 [w]\right).
\ee
This is in fact a general result.\footnote{We point out that to derive this formula, we have assumed the (normalised) rod directions have components $\ell_1 [\phi]=\ell_3 [\phi]=1$, which are needed to ensure that the orbits of $\frac{\partial}{\partial \phi}$ are identified with standard period $2\pi$ in the reduced Kaluza--Klein theory. This formula can be easily generalised to cases when $\ell_{1,3} [\phi]\ne 1$.} We remark that a similar result is known for four-dimensional magnetically charged Kaluza--Klein black holes \cite{Gibbons:1985,Rasheed:1995zv}. In this case, these black-hole solutions can be naturally lifted to five dimensions, and their magnetic charges then become NUT charges (modulo signs and factors of 2). It turns out that the NUT charges are simply encoded in the rod structure of these five-dimensional solutions \cite{Chen:2010zu,Chen:2010ih}. In fact, along this analogy, the dipole black ring considered here can be thought of as a string of magnetically charged Kaluza--Klein black holes, bent into a circular shape (parameterised by $\psi$).

The magnetic dipole potential $\Phi_{\rm m}$, first defined by Emparan for his $S^1$-rotating dipole black ring \cite{Emparan:2004wy}, now has a non-trivial generalisation when $S^2$ rotation is turned on \cite{Copsey:2005se}:
\be
\Phi_{\rm m}=-\frac{\pi}{2G_5}(B_{t\psi}+\Omega_{\phi}B_{\phi\psi})\big|_{\rm H}\,,
\ee
where the subscript ${\rm H}$ indicates that the evaluation is taken at the event horizon. Since $B=0$ at infinity, this potential should be understood as the difference between the potential at infinity and that on the horizon. Using the explicit expressions (\ref{B_field_components}), we find that the potential is a constant:
\be
\Phi_{\rm m}=\frac{\pi\varkappa \sqrt{2(a-c)\Psi} }{2G_5\sqrt{a\Phi}}\,.
\ee
Thus, if we define the temperature and entropy of the black hole as $T=\frac{\kappa}{2\pi}$ and $S=\frac{A_{\rm H}}{4G_5}$ respectively, it is straightforward to verify that the Smarr formula
\be
\frac{2}{3}\,M=TS+\Omega_{\psi}J_{\psi}+\Omega_{\phi} J_{\phi}+\frac{1}{3}\,\Phi_{\rm m}\mathcal{Q}\,,
\ee
and the first law of black-hole thermodynamics
\be
\dif M=T\,\dif S+\Omega_{\psi}\,\dif J_{\psi}+\Omega_{\phi}\,\dif J_{\phi}+\Phi_{\rm m}\,\dif \mathcal{Q}\,,
\ee
both hold for this dipole black ring solution.

Note that $g_{t\phi}$ vanishes along the two axes $x=\pm1$. This ensures that Dirac--Misner singularities are absent in the space-time. The absence of such singularities can in fact be seen from the rod structure in Fig.~\ref{Final rod structure}, from the fact that the directions of Rods 1 and 3 do not have time components \cite{Chen:2010zu,Chen:2010ih}. There are also no conical singularities in the space-time. From the rod-structure viewpoint, this follows from the fact that the directions of Rods 1 and 3 are the same.

It has been checked numerically that both $H(x,y)>0$ and $K(x,y)>0$ everywhere on and outside the horizon. Since the curvature invariants have denominators that are proportional to some positive powers of $H(x,y)$ and $K(x,y)$, their positivity will imply that there are no curvature singularities in this region. It should also be checked if this region contains closed time-like curves (CTCs). Despite an extensive numerical search, no CTCs were found anywhere in this region. Thus, it would appear that the space-time on and outside the event horizon is regular and well behaved.

\newsection{Phase space structure}

We can fix the overall scale of the balanced doubly rotating dipole black ring by fixing its mass $M$. The solution is then characterised by reduced dimensionless quantities obtained by dividing out an appropriate power of $M$ or $G_5M$, which in this case are \cite{Emparan:2004wy}
\ba
j_\psi^2&\equiv&\frac{27\pi}{32G_5}\,\frac{J_\psi^2}{M^3}=\frac{27a(a+c)\Phi^2\left[(1+c)\Phi+2bc(1-a)\right]^2}{16(1+b)\left[(a+c)\Phi+a(1-b+c+bc)\right]^3}\,,\cr
j_\phi^2&\equiv&\frac{27\pi}{32G_5}\,\frac{J_\phi^2}{M^3}=\frac{27abc^2(a+c)(1-a^2)\Phi^2}{4(1+b)\left[(a+c)\Phi+a(1-b+c+bc)\right]^3}\,,\cr
q^2&\equiv&\frac{\mathcal{Q}^2}{{G_5M}}={\frac{2a(1+b)(a-c)}{\pi\left[(a+c)\Phi+a(1-b+c+bc)\right]}}\,,\cr
a_{\mathrm{H}}&\equiv&\frac{3}{16}\,\sqrt{\frac{3}{\pi}}\,\frac{A_{\mathrm{H}}}{(G_5M)^{\frac{3}{2}}}=\sqrt{\frac{6a(a+c)(1-a^2)}{1+b}}\,\frac{3c\Phi}{\left[(a+c)\Phi+a(1-b+c+bc)\right]^{\frac{3}{2}}}\,.
\ea
The phase space of this solution is therefore a four-dimensional one, parameterised by $(j_\psi, j_\phi, q, a_{\mathrm{H}})$. To study it, we will look at its various two-dimensional cross-sections; without loss of generality, we take $j_\psi,j_\phi,q\geq0$.

\begin{figure}[t]
\begin{center}
\includegraphics[scale=0.625, angle=0]{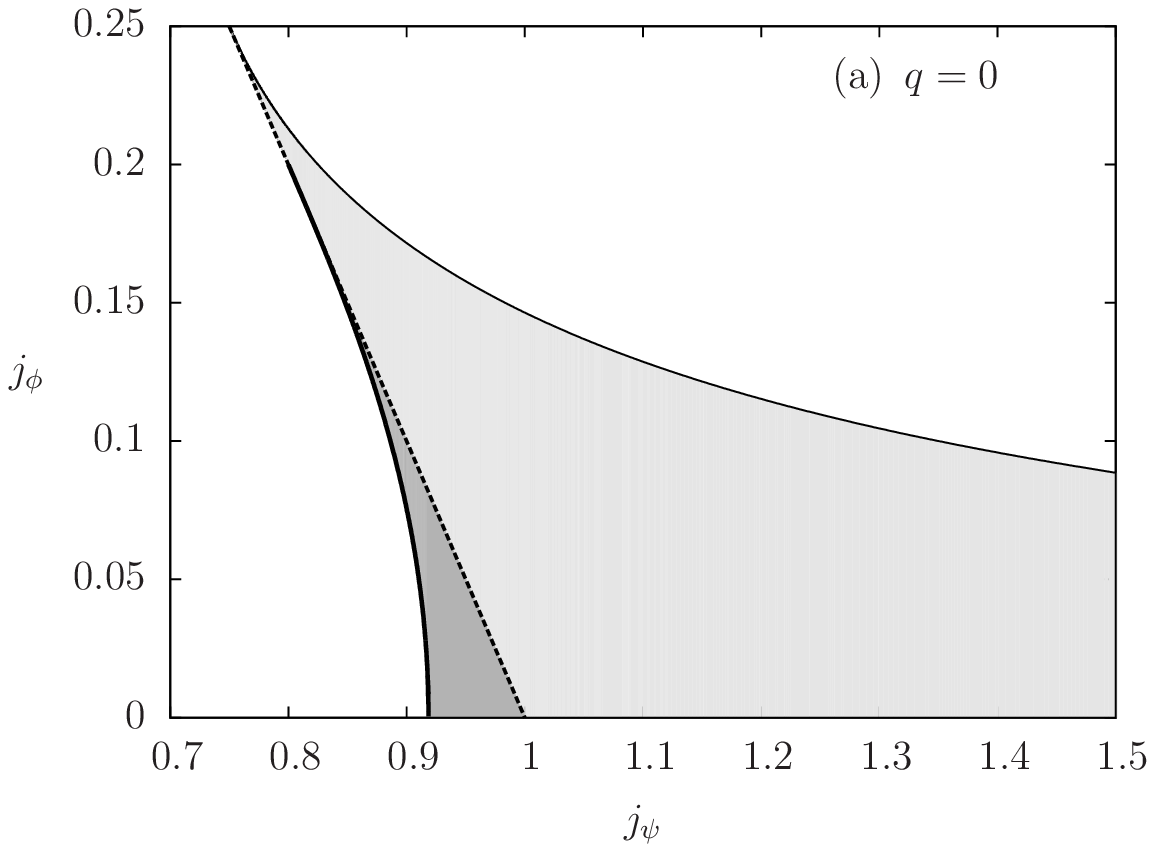}
\includegraphics[scale=0.625, angle=0]{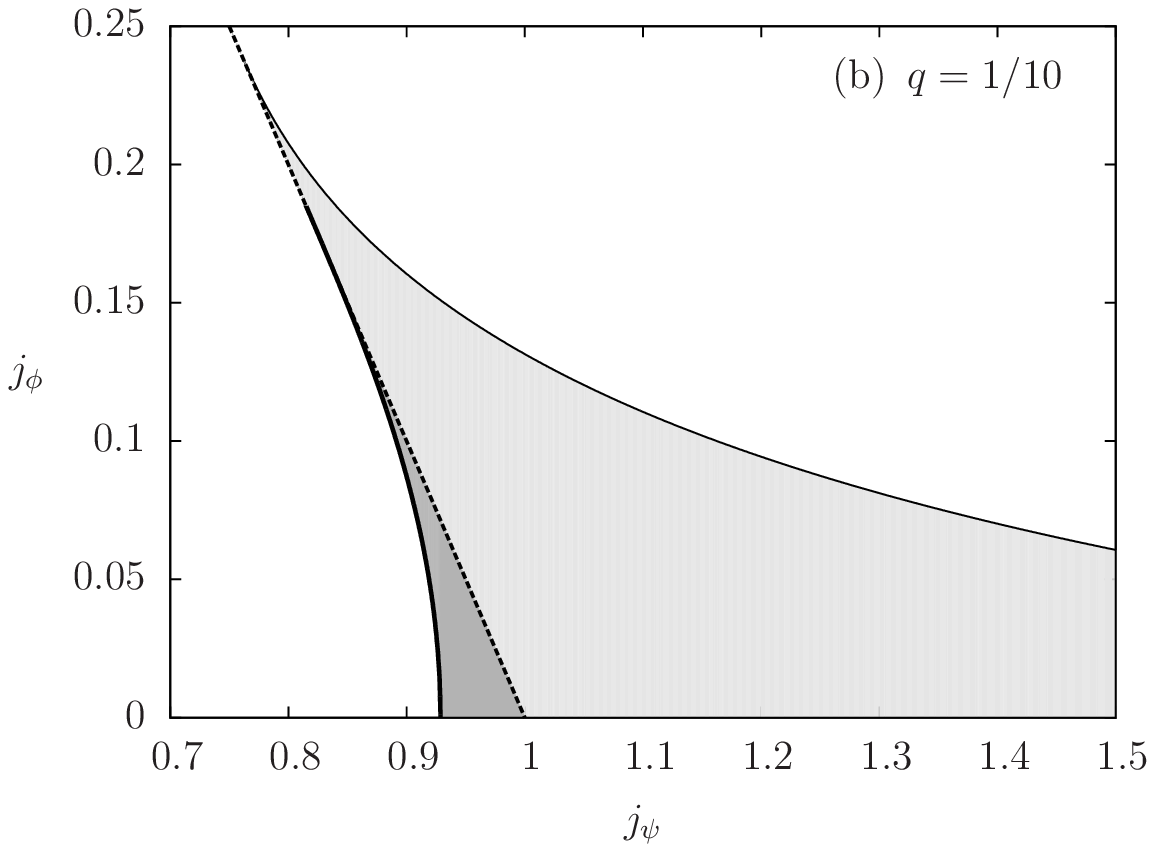}
\includegraphics[scale=0.625, angle=0]{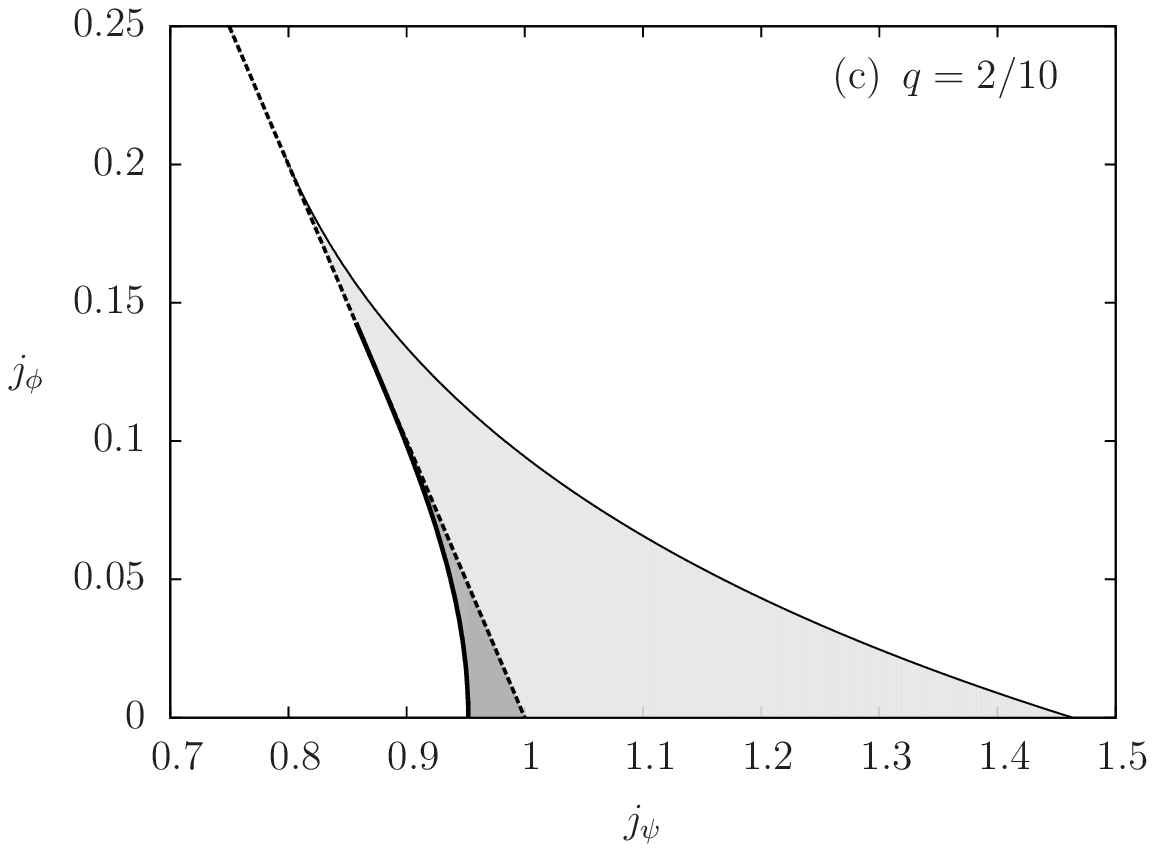}
\includegraphics[scale=0.625, angle=0]{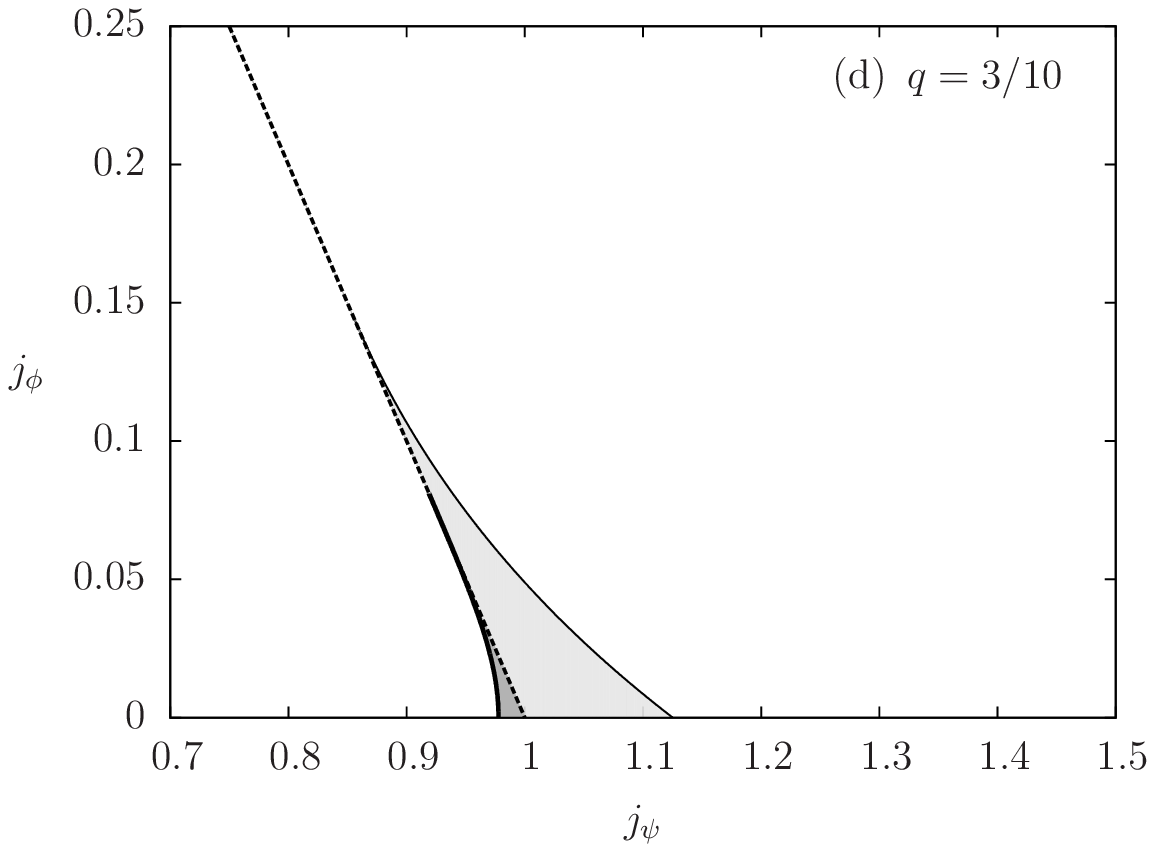}
\caption{$(j_\psi, j_\phi)$ phase diagrams for the respective values $q=0,\frac{1}{10},\frac{2}{10}, \frac{3}{10}$. In each case, the phase space is divided into two distinct regions by the dashed line. To the left of it is a dark-grey region bounded by the thick black curve and the $j_\phi=0$ axis. To the right is a light-grey region bounded by the thin black curve and the $j_\phi=0$ axis. 
\label{fig_j1_j2}}
\end{center}
\end{figure}

We begin by studying the $(j_\psi, j_\phi)$ phase diagram for fixed-$q$ values, as in Fig.~\ref{fig_j1_j2}. As a reference, we have also plotted the $q=0$ case, which was previously studied in \cite{Emparan:2008eg}. Note that in each case, the phase space is bounded by three curves (besides the $j_\phi=0$ axis). They are:

\begin{enumerate}
\item A {\it thin black curve\/}, corresponding to extremal doubly rotating dipole black rings. These black rings are obtained by maximising $j_\phi$ for a fixed value of $j_\psi$. This limit will be examined in more detail in Sec.~6.3.

\item A {\it thick black curve\/}, corresponding to non-extremal minimally $S^1$-rotating dipole black rings. These black rings are obtained by minimising $j_\psi$ for a fixed value of $j_\phi$, which determines $b$ in terms of $a$ and $c$ as
\be
b_{\mathrm{crit}}=\frac{1+a}{1-a}\,\frac{(1-a)(a-c)^2-3a^2(1+c)(1-a-c)}{(1+a)(a-c)^2-3a^2(1-c)(1+a+c)}\,.
\ee
It turns out that this curve only exists for sufficiently small $j_\phi$. For larger values of $j_\phi$, the minimal $j_\psi$ boundary will be replaced by the following curve.

\item A {\it dashed line\/}, given by $j_\psi+j_\phi=1$, corresponding to the so-called collapse limit of the black ring. In this limit, the black ring collapses to an extremal Myers--Perry black hole, and so the dashed line is actually not part of the black-ring phase space. This limit will be examined in more detail in Sec.~6.4.
\end{enumerate}

The regions of the phase space where black rings exist are shaded in grey in Fig.~\ref{fig_j1_j2}. Note that in each case, there are two distinct regions separated by the dashed line. To its left (where $j_\psi+j_\phi<1$) is a dark-grey region bounded by the thick black curve and the $j_\phi=0$ axis. To its right (where $j_\psi+j_\phi>1$) is a light-grey region bounded by the thin black curve and the $j_\phi=0$ axis; in this region, the black rings have the property of being ``thin''. By contrast, both ``thick'' and ``thin'' black rings co-exist in the dark-grey region. (This fact cannot be seen from Fig.~\ref{fig_j1_j2}, and will only be apparent when we come to Fig.~\ref{fig_j12_aH} below.)

From the phase diagrams in Fig.~\ref{fig_j1_j2}, it is clear that the upper bound for $j_\phi$ is reached when the thin black curve meets the dashed line. This corresponds to the collapse limit of the extremal doubly rotating dipole black ring. At this point, it can be checked that
\be
(j_\phi)_{\rm max}=\frac{1}{4}-\frac{3\pi}{8}\,q^2.
\ee
Note that this point also gives the lower bound for $j_\psi$ for fixed $q$:
\be
(j_\psi)_{\rm min}=\frac{3}{4}+\frac{3\pi}{8}\,q^2.
\ee

When $q=0$, the allowed range for $j_\psi$ extends to infinity. However, when $q\neq0$, an upper bound for $j_\psi$ appears, given by
\be
(j_\psi)_{\rm max}=\sqrt{\frac{27}{2048\pi}}\,\frac{(2+\pi q^2)^2}{q}\,.
\ee
This upper bound is achieved by the extremal singly rotating dipole ring, which is actually singular \cite{Emparan:2004wy}.
As $q$ is increased from 0, this upper bound starts to decrease. At the same time, the upper bound for $j_\phi$ for fixed $j_\psi$ (the thin black curve) decreases. This leads to an overall shrinking of the area of the light-grey region.
It turns out that the dark-grey region also shrinks in area. This can be seen from the fact that as $q$ is increased, the lower bound for $j_\psi$ for fixed $j_\phi$ (the thick black curve) is increased, together with a decreased value of its upper bound (where the thick black curve meets the dashed line).

\begin{figure}[t]
\begin{center}
\includegraphics[scale=0.625, angle=0]{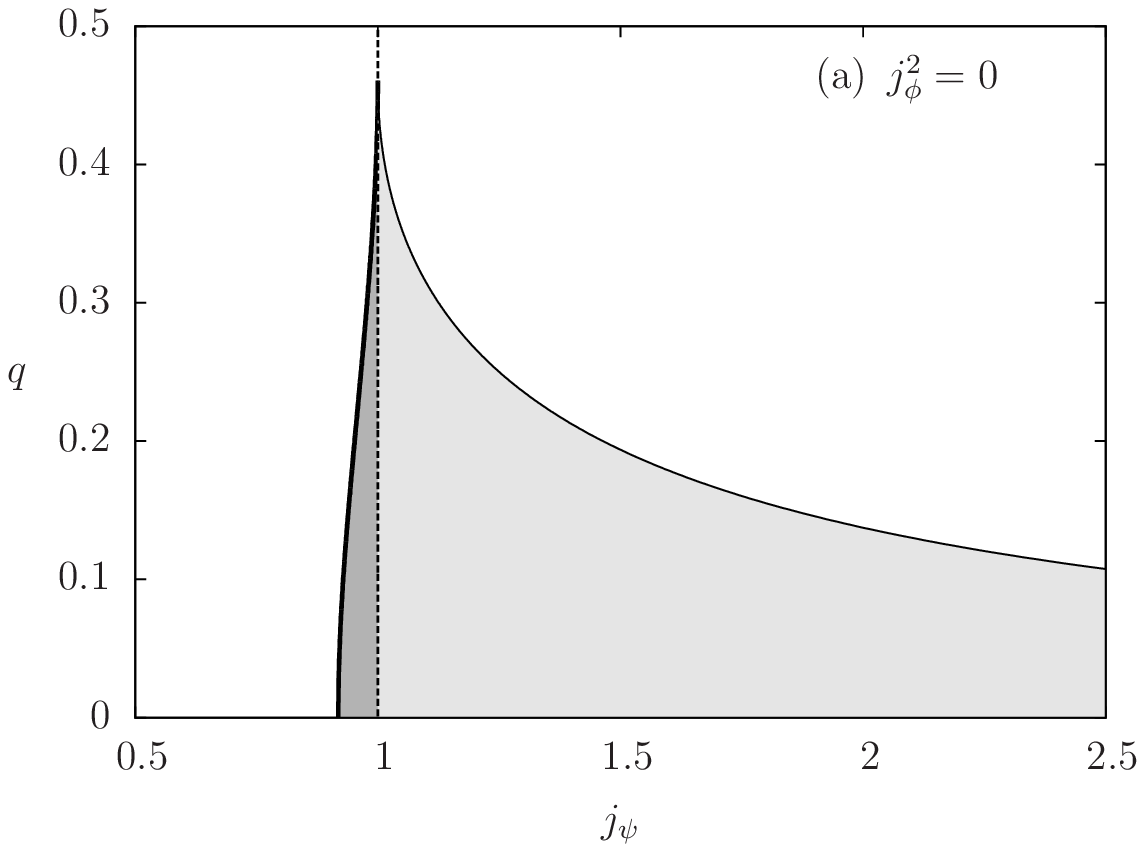}
\includegraphics[scale=0.625, angle=0]{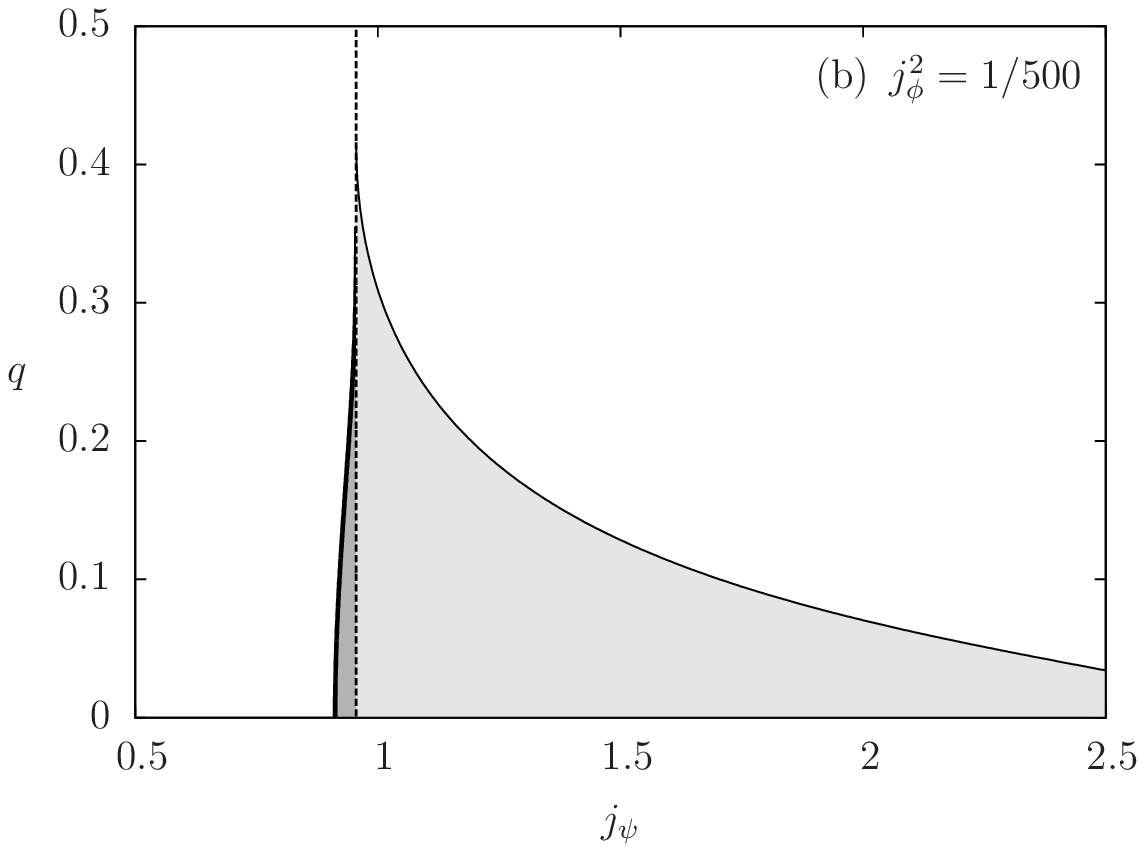}
\includegraphics[scale=0.625, angle=0]{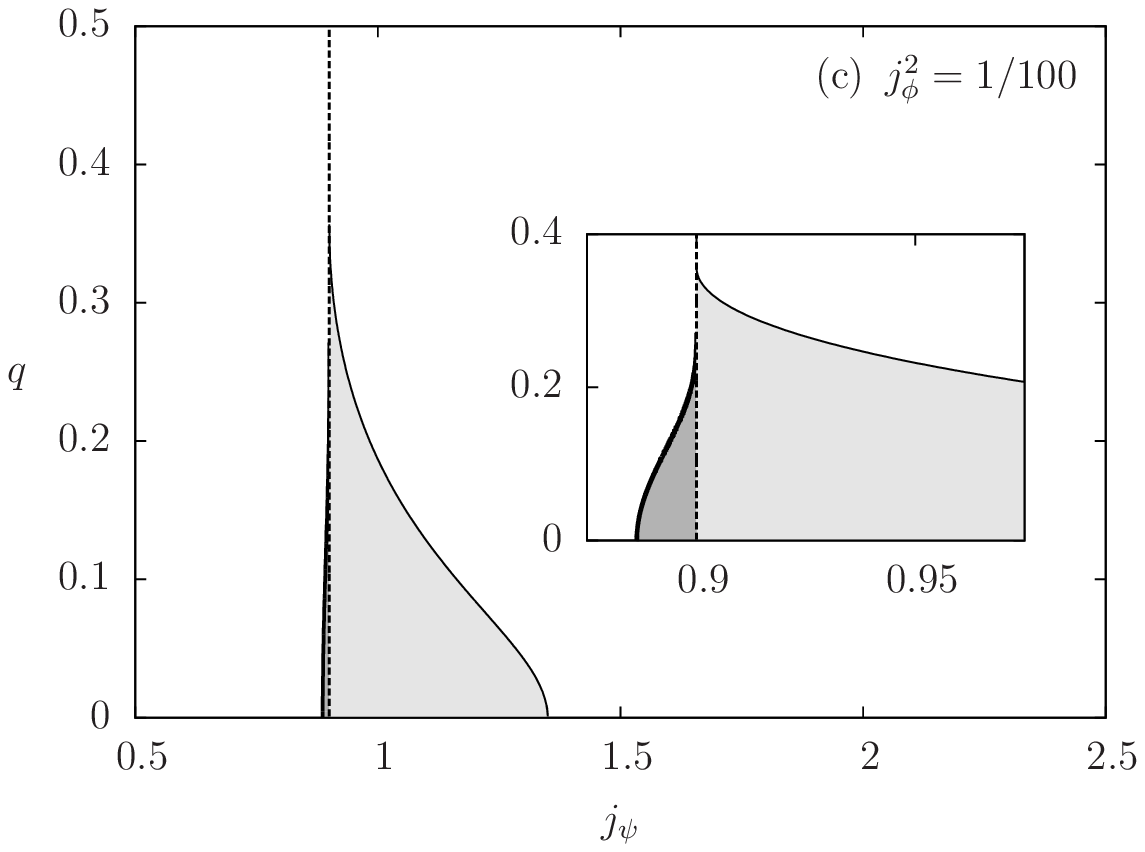}
\includegraphics[scale=0.625, angle=0]{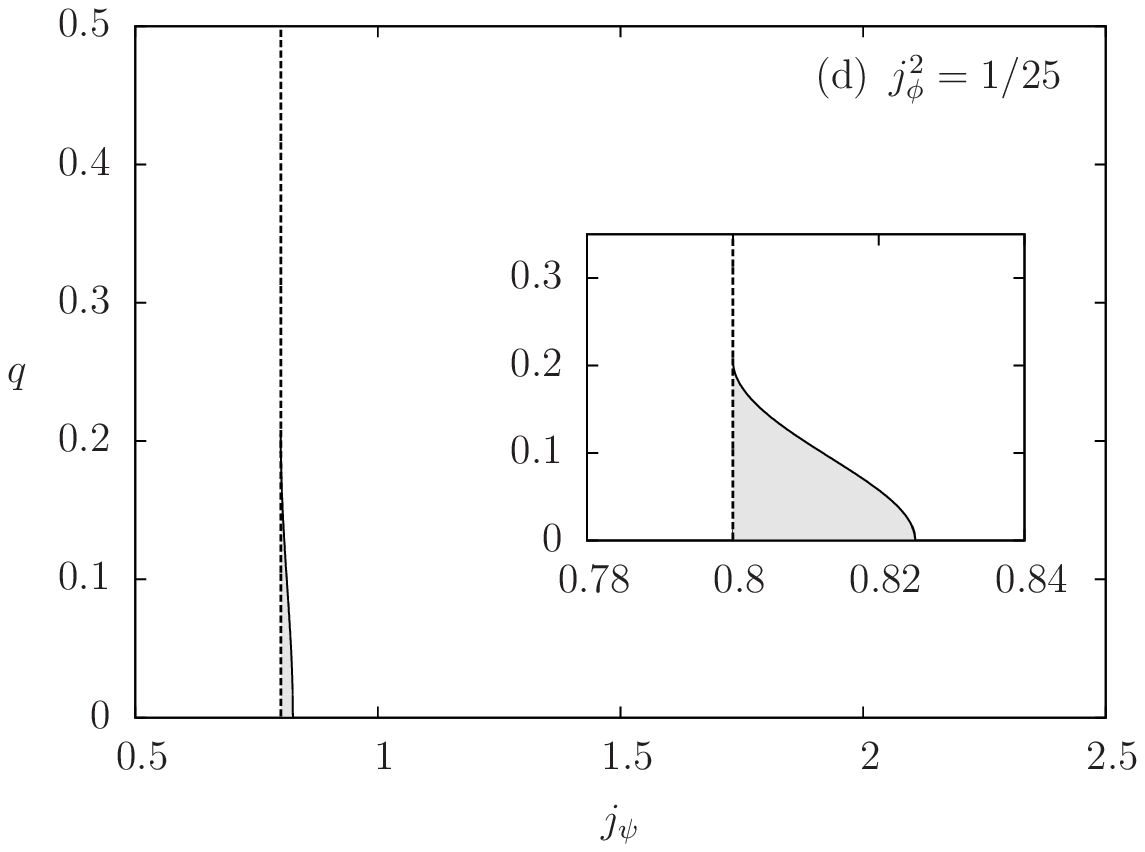}
\caption{$(j_\psi, q)$ phase diagrams for the respective values $j_\phi^2=0,\frac{1}{500},\frac{1}{100},\frac{1}{25}$. For the latter two cases, the region near $q=0$ has been magnified in the insets for clarity. Note that the dark-grey region has disappeared in the last case.
\label{fig_j1_q}}
\end{center}
\end{figure}

We next turn to the $(j_\psi, q)$ phase diagram for fixed-$j_\phi^2$ values, as in Fig.~\ref{fig_j1_q}. In each of these diagrams, the phase space is bounded by the same three boundary curves as in the $(j_\psi, j_\phi)$ phase diagram. Again, the dashed line divides the phase space into two distinct regions: a dark-grey region on its left (where $j_\psi+j_\phi<1$) bounded by the thick black curve and the $q=0$ axis, and a light-grey region on its right (where $j_\psi+j_\phi>1$) bounded by the thin black curve and the $q=0$ axis. For $j_\phi\geq\frac{1}{5}$, the thick black curve ceases to exist and the dark-grey region disappears.

From the phase diagrams in Fig.~\ref{fig_j1_q}, it is clear that the upper bound for $q$ is reached when the thin black curve meets the dashed line. This again corresponds to the collapse limit of the extremal doubly rotating dipole black ring, except that $j_\phi$ is fixed in this case. It can be checked that
\be
q_{\rm max}=\sqrt{\frac{2(1-4j_\phi)}{3\pi}}\,.
\ee
For fixed values of $j_\psi$, the upper bound for $q$ is given by the thin black curve in the light-grey region, and by the thick black curve in the dark-grey region (if it exists). As  $j_\phi$ is increased, this upper bound decreases, and both the light- and dark-grey regions shrink in area.

\begin{figure}[t]
\begin{center}
\includegraphics[scale=0.625, angle=0]{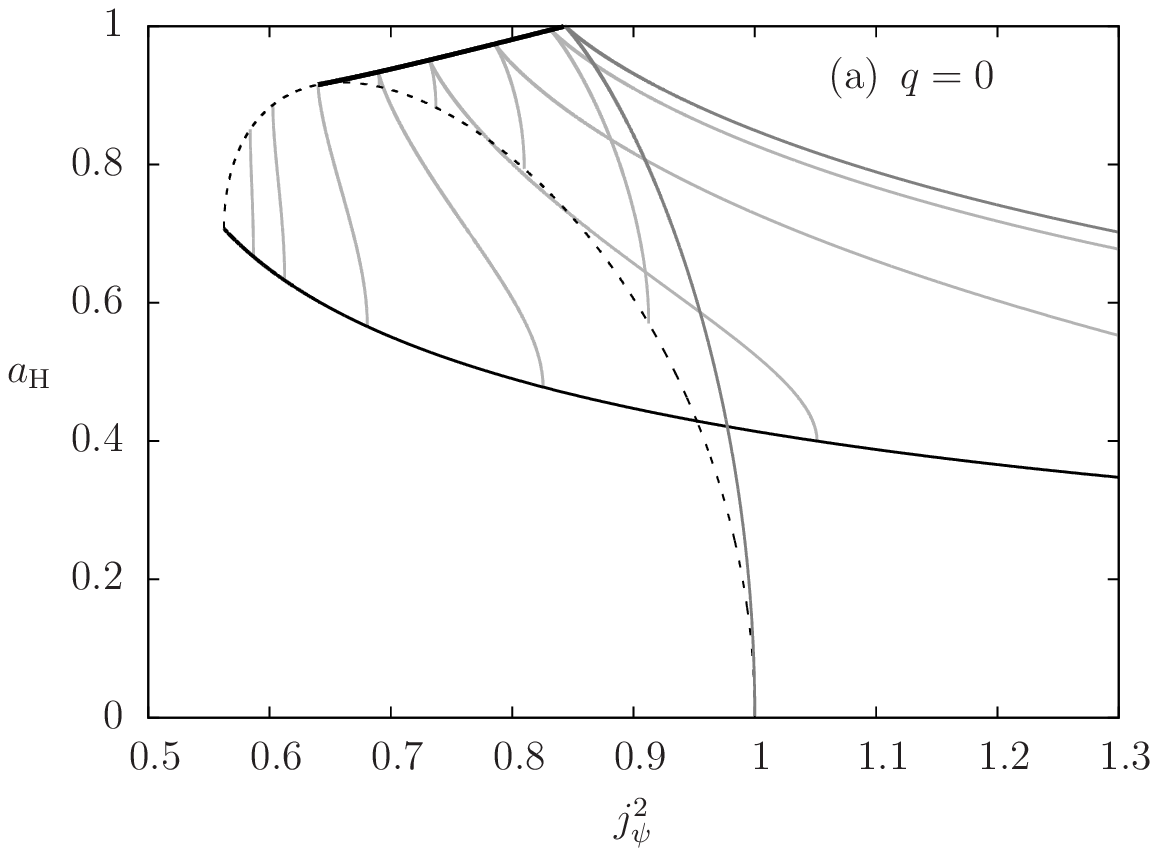}
\includegraphics[scale=0.625, angle=0]{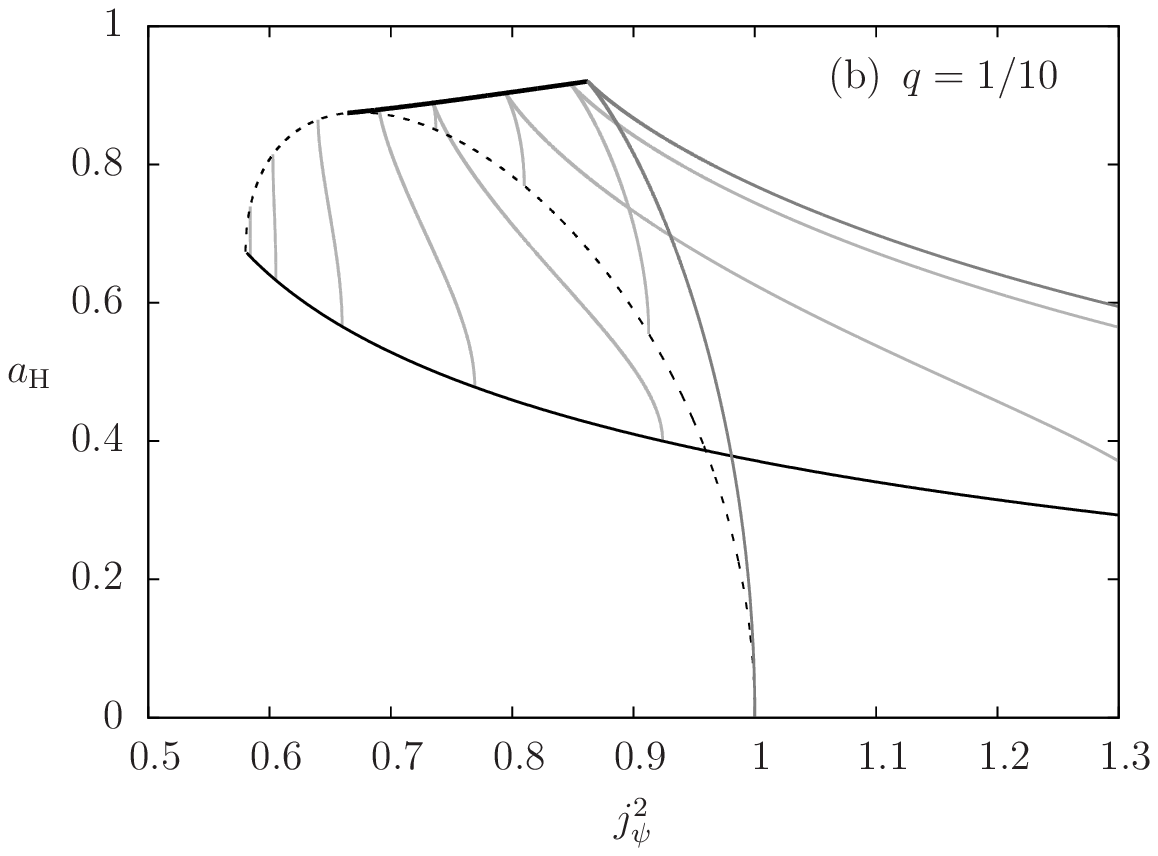}
\includegraphics[scale=0.625, angle=0]{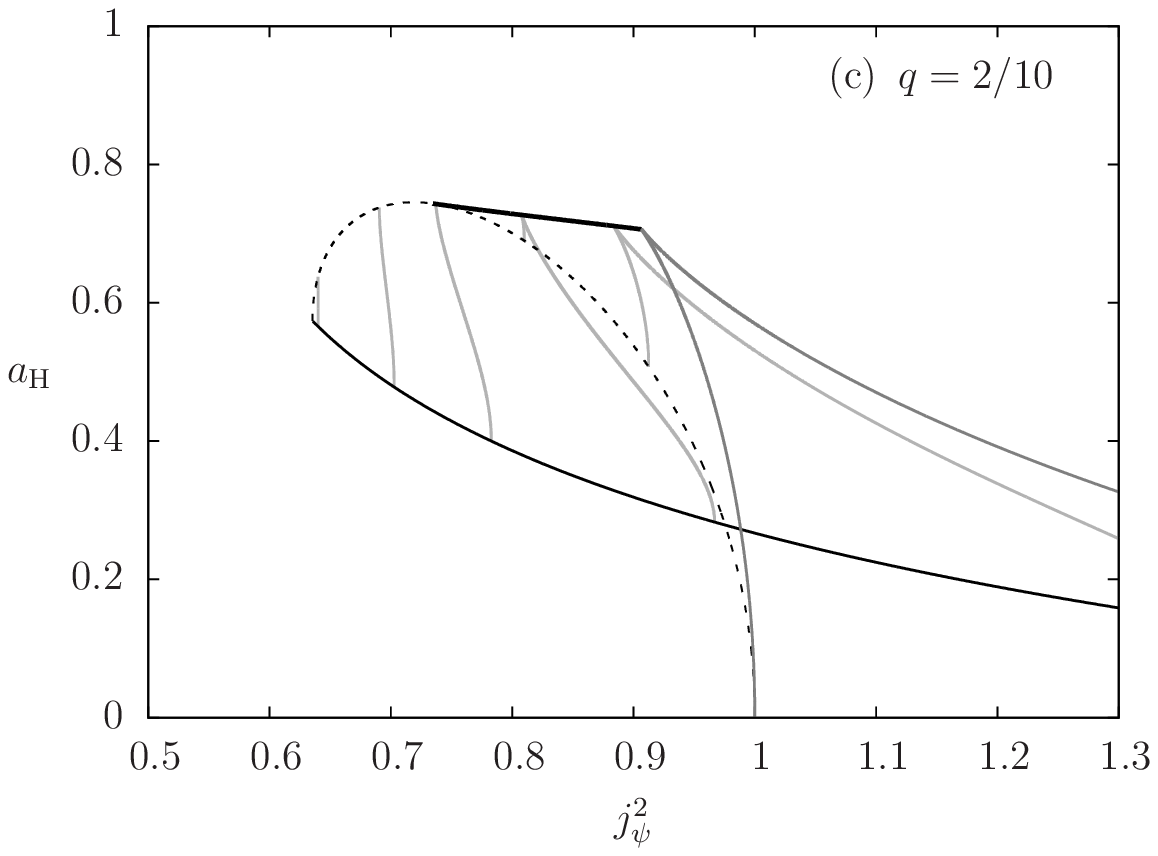}
\includegraphics[scale=0.625, angle=0]{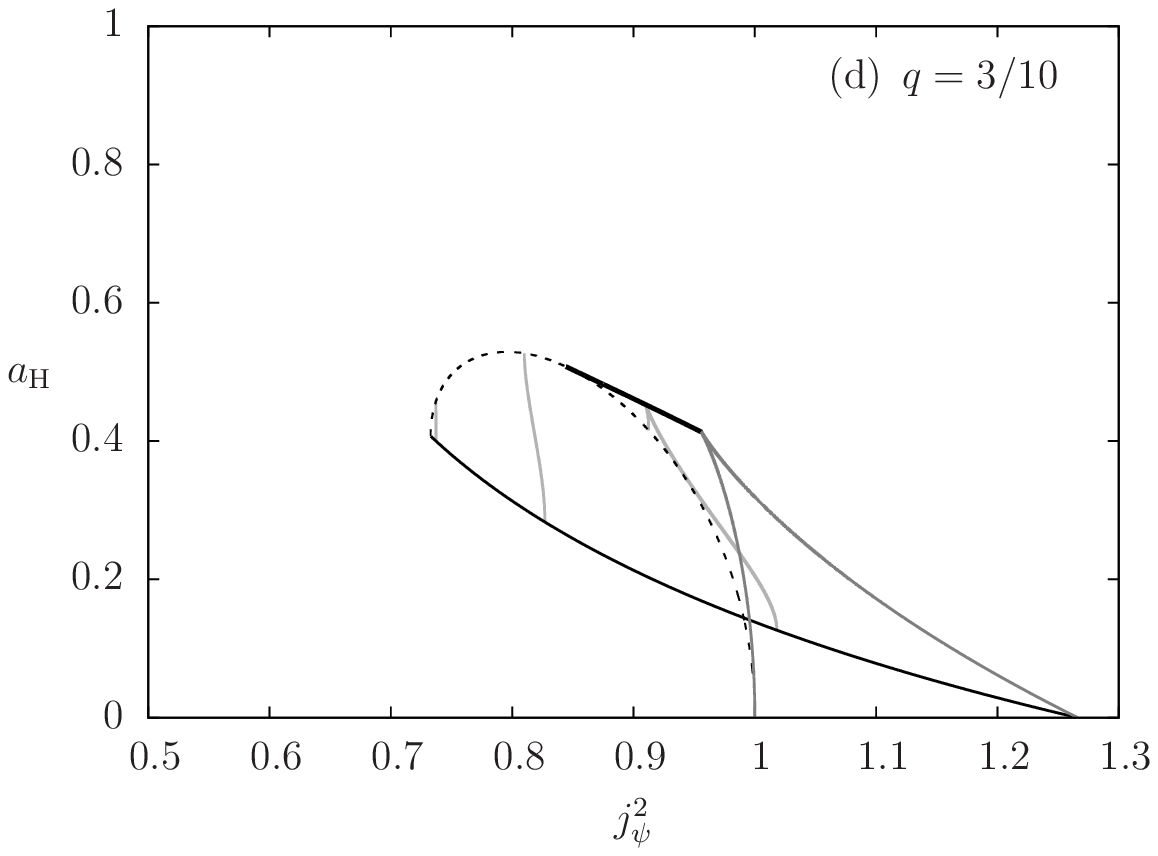}
\caption{$(j_\psi^2, a_{\mathrm{H}})$ phase diagrams for the respective values $q=0,\frac{1}{10},\frac{2}{10}, \frac{3}{10}$. In each case, the dark-grey curve shows the phase of the singly rotating black ring. The light-grey curves are branches of constant $j_\phi$; the particular values shown are (from right to left, and if the value of $j_\phi^2$ is allowed for the corresponding value of $q$) $j_\phi^2=\frac{1}{500},\frac{1}{100},\frac{1}{50},\frac{1}{35},\frac{1}{25},\frac{1}{20},\frac{1}{18}$. 
\label{fig_j12_aH}}
\end{center}
\end{figure}

Finally, we turn to the $(j_\psi^2, a_{\mathrm{H}})$ phase diagram for fixed-$q$ values, as in Fig.~\ref{fig_j12_aH}. In each of these diagrams, we have plotted the three boundary curves of the $(j_\psi, j_\phi)$ phase diagram. In addition, there are a number of grey curves plotted. The dark-grey curve represents the phase of the singly rotating black ring \cite{Emparan:2004wy}, while the light-grey curves are branches of constant $j_\phi>0$. 

The reference case $q=0$ was previously studied in \cite{Elvang:2007hs}. Each of the light-grey constant-$j_\phi$ curves starts at the thin black curve and ends up at the dashed curve. In the process, there is a cusp formed at the thick black curve. It is this cusp that separates the thin-ring branch from the thick-ring branch. The totality of all these light-grey curves then gives the phase space of allowed black rings. On the other hand, the space traced out by all the thick-ring branches (the part of each light-grey curve starting from the thick black curve and ending at the dashed curve) is the subset of this total phase space corresponding to the dark-grey region of Fig.~\ref{fig_j1_j2}(a). This is the region in which both thin and thick rings can co-exist. Note also that there are light-grey curves without a cusp (and therefore a thick-ring branch); these start from the thin black curve and end up directly at the dashed curve. Such curves only exist for sufficiently large values of $j_\phi$. 

As $q$ is increased from 0, it is evident that the upper bound for the horizon area $a_{\rm H}$ decreases. At the same time, the allowed range for $j_{\psi}$ becomes narrower: as we have seen above, $(j_{\psi})_{\rm min}$ increases while $(j_{\psi})_{\rm max}$ decreases as $q$ is increased. In particular, let us focus on the minimally $S^1$-rotating dipole black rings (the thick black curves in Fig.~\ref{fig_j12_aH}). Observe that for fixed $j_\phi$, the minimally rotating dipole black ring has a smaller $a_{\rm H}$ and a larger $j_\psi$ than the corresponding $q=0$ black ring. This former is to be expected: in general, the area of a black hole decreases when charge is added to it. On the other hand, the latter follows from the fact that there is now an enhanced attraction between opposite sides of the ring, due to their opposite charges. A larger centrifugal force is thus needed to balance the ring. We note that similar observations have been made in the $j_\phi=0$ case \cite{Emparan:2004wy}. 

\bigbreak
\newsection{Various limits}

\newsubsection{Singly rotating dipole black ring}

As mentioned above, the parameter $b$ governs the $S^2$ rotation of the dipole black ring. If we set $b=0$,
the $S^2$ rotation vanishes, and the resulting black ring rotates only in the $S^1$ direction. This is the (balanced) dipole black ring discovered by Emparan \cite{Emparan:2004wy}, for the special case $\alpha=2\sqrt{2/3}$ corresponding to Kaluza--Klein theory. The metric (\ref{solution5D}) then becomes
\ba
\dif s^2&=&-\frac{\tilde{F}(y)}{\tilde{F}(x)}\,\bigg[\frac{\tilde{H}(x)}{\tilde{H}(y)}\bigg]^{\frac{1}{3}}\left(\dif t+\tilde{\omega}_1\,\dif\psi\right)^2+\frac{2\varkappa^2}{(x-y)^2}\tilde{F}(x)\left[\tilde{H}(x)\tilde{H}(y)^2\right]^{\frac{1}{3}}\,\cr
&&\times\left\{-\frac{G(y)\,\dif\psi^2}{\tilde{F}(y)\tilde{H}(y)}+\frac{G(x)\,\dif\phi^2}{\tilde{F}(x)\tilde{H}(x)}+\frac{1}{1-a^2}\bigg[\frac{\dif x^2}{G(x)}-\frac{\dif y^2}{G(y)}\bigg]\right\},
\ea
where
\be
\tilde{\omega}_1=-\sqrt{\frac{2a(1+a)(a+c)}{1-a}}\,\frac{\varkappa(1+c)(1+y)}{\tilde{F}(y)}\,,
\ee
and the functions $\tilde{F}$ and $\tilde{H}$ are given by
\be
\tilde{F}(x)=1+ac+(a+c)x\,,\qquad\tilde{H}(x)=1-ac-(a-c)x\,.
\ee
The gauge potential $\tilde{A}$ and dilaton field $\tilde{\varphi}$ are given by
\be
\tilde{A}=\sqrt{\frac{2a(1-a)(a-c)}{1+a}}\,\frac{\varkappa(1+c)(1+x)}{\tilde{H}(x)}\,\dif\phi\,,\qquad \me^{-\tilde{\varphi}}=\bigg[\frac{\tilde{H}(x)}{\tilde{H}(y)}\bigg]^{\sqrt{\frac{2}{3}}}.
\ee

To recover the exact form of the solution given in \cite{Emparan:2004wy}, we need to define the parameters
\ba
a=\frac{\mu+\nu}{1+\mu\nu}\,,\qquad c=\nu\,,\qquad \varkappa=R\,\sqrt{\frac{1-\mu^2}{2(1+\nu^2+2\mu\nu)}}\,,
\ea
and the coordinates
\be
(\psi,\phi)=\sqrt{\frac{1+\nu^2+2\mu\nu}{1-\mu^2}}\,(-\tilde{\psi},\tilde{\phi})\,,
\ee
where the coordinates $\tilde{\psi}$ and $\tilde{\phi}$ are identified with the coordinates $\psi$ and $\phi$ of \cite{Emparan:2004wy}, respectively.

\newsubsection{Pomeransky--Sen'kov black ring}

The dipole charge $\mathcal{Q}$ of the solution (\ref{solution5D})--(\ref{phi_soln}) vanishes in the limit $a=c$, in which case the Maxwell field vanishes, and the dilaton field becomes constant and decouples from the system. The metric in this limit reduces to that of the Pomeransky--Sen'kov black ring \cite{Pomeransky:2006bd}. The explicit form of the metric in the current coordinates is
\ba
\dif s^2&=&-\frac{\tilde{H}(y,x)}{\tilde{H}(x,y)}\,\left(\dif t+\tilde{\omega}_1\,\dif\psi+\tilde{\omega}_2\,\dif\phi\right)^2+\frac{2\varkappa^2\tilde{H}(x,y)}{(x-y)^2}\,\bigg\{\frac{\tilde{F}(x,y)\left(\dif\psi+\tilde{\omega}_3\,\dif\phi\right)^2}{\tilde{H}(x,y)\,\tilde{H}(y,x)}\cr
&&-\frac{G(x)G(y)\,\dif\phi^2}{\tilde{F}(x,y)}+\frac{1}{\tilde\Phi\tilde\Psi}\bigg[\frac{\dif x^2}{G(x)}-\frac{\dif y^2}{G(y)}\bigg]\bigg\}\,,
\ea
where
\ba
\tilde{\omega}_1&=&\frac{2\varkappa c(1+b)(1+y)}{\sqrt{\tilde\Phi\tilde\Psi}\,\tilde{H}(y,x)}\,\Big\{b(1+cx)(1+cy)\left[b(c-1)(1+cx)+2+2c\right]\cr
&&-(1-c^2)\left[bc(1+c+y+cx^2y)+(1+c)^2\right]\Big\}\,,\cr
\tilde{\omega}_2&=&\sqrt{\frac{b(1-c^2)}{\tilde\Phi\tilde\Psi}}\,\frac{2\varkappa c^2(1+b)(1-x^2)}{\tilde{H}(y,x)}\,\left[(1+cy)(c+bc+by)-c-y\right],\cr
\tilde{\omega}_3&=&\frac{\sqrt{b(1-c^2)}}{\tilde\Phi\tilde\Psi}\,\frac{c^2(1+b)(x-y)(1-x^2)(1-y^2)}{\tilde{F}(x,y)}\,\cr
&&\times\left[b(1+cx)(1+cy)(1-b-c^2-bc^2)-(1-c^2)(1-b+c^2+bc^2)\right],
\ea
and the functions $\tilde{H}$ and $\tilde{F}$ are given by
\ba
\tilde{H}(x,y)&=&-c^2(1+b)\left[bx^2(1+cy)^2+(c+x)^2\right]+\left[b(1+cy)-1-cx\right]^2+bc^2(1-xy)^2,\cr
\tilde{F}(x,y)&=&\frac{1-y^2}{\tilde\Phi\tilde\Psi}\,\bigg\{bcG(x)\Big\{c(y^2-1)\left[c^2(1+b)-b+1\right]^2-4c^2y(1-b^2)(1+cy)\Big\}\cr
&&-(1+cy)\Big\{c^2(1+b)^2\left[c^2(c+x+bx+bcx^2)^2-(c+x-bx-bcx^2)^2\right]\cr
&&-(1-b)^2(1+cx)^2\left[c^2(1+b)^2-(1-b)^2\right]\Big\}\bigg\}\,.
\ea
Here, $\tilde\Phi=1-b+c+bc$ and $\tilde\Psi=1-b-c-bc$.

To obtain this black ring in the form given in \cite{Chen:2011jb} (in turn closely related to the original form of \cite{Pomeransky:2006bd}), we need to perform the following parameter redefinitions and coordinate transformations:
\be
b=\frac{\nu(1-\mu^2)}{\mu(1-\nu^2)}\,,\qquad c=\frac{\mu-\nu}{1-\mu\nu}\,,\qquad x=\frac{\tilde{x}+\nu}{1+\nu \tilde{x}}\,,\qquad y=\frac{\tilde{y}+\nu}{1+\nu \tilde{y}}\,,
\ee
where the coordinates $\tilde{x}$ and $\tilde{y}$ are identified with the coordinates $x$ and $y$ of \cite{Chen:2011jb}, respectively.

\newsubsection{Extremal doubly rotating dipole black ring}

The extremal limit of the solution (\ref{solution5D})--(\ref{phi_soln}) in the present form is a bit subtle: it corresponds to $a,c\rightarrow 0$ and $b\rightarrow 1$. More precisely, we first define
\be
a=\frac{c}{2\alpha}\,,\qquad b=1-\frac{c}{\beta}\,,
\ee
where $\alpha$ and $\beta$ are new parameters satisfying $0<\beta<\alpha\le1/2$, and then take the limit $c\rightarrow 0$. 
The explicit form of the metric obtained is
\ba
\label{extremal_limit}
\dif s^2&=&-\left[\frac{\tilde{H}(y,x)^3}{\tilde{K}(x,y)^2\tilde{H}(x,y)}\right]^{\frac{1}{3}}\left(\dif t+\tilde{\omega}_1\,\dif\psi+\tilde{\omega}_2\,\dif\phi\right)^2+\frac{2\varkappa^2}{(x-y)^2}\left[\tilde{K}(x,y)\tilde{H}(x,y)^2\right]^{\frac{1}{3}}\cr
&&\times\bigg\{\frac{\tilde{F}(x,y)\left(\dif\psi+\tilde{\omega}_3\,\dif\phi\right)^2}{\tilde{H}(x,y)\tilde{H}(y,x)}-\frac{\tilde{G}(x)\tilde{G}(y)\,\dif\phi^2}{\tilde{F}(x,y)}+\frac{1}{\alpha^2-\beta^2}\bigg[\frac{\dif x^2}{\tilde{G}(x)}-\frac{\dif y^2}{\tilde{G}(y)}\bigg]\bigg\}\,,\qquad~
\ea
where
\ba
\tilde{\omega}_1&=&-\sqrt{\frac{2(1+2\alpha)}{\alpha^2-\beta^2}}\,\frac{\varkappa\beta(1+y)}{\tilde{H}(y,x)}\big\{(\alpha+\beta)(\alpha-\beta x)+\alpha\beta[\alpha(1-y)+\beta(x^2-y)]\cr
&&+\alpha^2\beta^2(1-x^2)(1-y)\big\}\,,\cr
\tilde{\omega}_2&=&\sqrt{\frac{2(1+2\alpha)}{\alpha^2-\beta^2}}\,\frac{\varkappa\alpha\beta^2(1-x^2)}{\tilde{H}(y,x)}\,(\alpha\beta y^2-\alpha y-\alpha\beta+\beta)\,,\cr
\tilde{\omega}_3&=&\frac{\alpha\beta^2}{\alpha^2-\beta^2}\,\frac{(x-y)(1-x^2)(1-y^2)}{\tilde{F}(x,y)}\left[\alpha^2\beta(x+y)-\alpha^2-\beta^2\right]\,,
\ea
and the functions $\tilde{G}$, $\tilde{H}$, $\tilde K$  and $\tilde{F}$ are given by
\ba
\tilde{G}(x)&=&1-x^2,\cr
\tilde{H}(x,y)&=&(\alpha+\beta x)(\alpha-\beta y)-\alpha\beta^2(x^2-y^2)+\alpha^2\beta^2(1-x^2)(1-y^2)\,,\cr
\tilde{K}(x,y)&=&\alpha^2[1+\beta(x-y)]^2-\beta^2x^2+\alpha^2\beta^2(1-xy)^2,\cr
\tilde{F}(x,y)&=&\frac{1-y^2}{\alpha^2-\beta^2}\,\Big\{\alpha^2\beta^2\left\{(1+x)^2[1+\beta(1-x)]^2-2\beta(1-x^2)(1+y)\right\}\cr
&&-\alpha^4\beta^2(1-x^2)(1-y^2)-(\alpha^2+\beta^2x)^2\Big\}\,.
\ea
The gauge potential is given by $\tilde{A}=\tilde{A}_t\,\dif t+\tilde{A}_\psi\,\dif\psi+\tilde{A}_\phi\,\dif\phi$, where
\ba
\tilde{A}_t&=&-\sqrt{1-4\alpha^2}\,\frac{\alpha\beta^2(x-y)(1-xy)}{\tilde{K}(x,y)}\,,\cr
\tilde{A}_\psi&=&-\sqrt{\frac{2(1-2\alpha)}{\alpha^2-\beta^2}}\,\frac{\varkappa\alpha\beta^2(1+y)}{\tilde{K}(x,y)}\,\big\{(\alpha+\beta)x(1-y)+\beta(1-x)^2[1+\alpha(1+y)]\big\}\,,\cr
\tilde{A}_\phi&=&\sqrt{\frac{2(1-2\alpha)}{\alpha^2-\beta^2}}\,\frac{\varkappa\beta(1+x)}{\tilde{K}(x,y)}\,\big\{(\alpha-\beta)(\alpha+\beta x)+\alpha\beta[\alpha(1+x-2y)+\beta y(1-x)]\cr
&&+\alpha^2\beta^2(1+x)(1-y)^2\big\}\,,
\ea
and the dilaton field is given by 
\be
\me^{-\varphi}=\left[\frac{\tilde{K}(x,y)}{\tilde{H}(x,y)}\right]^{\sqrt{\frac{2}{3}}}.
\ee

In this limit, the outer horizon recedes to $y=-\infty$ and coincides with the inner one. The horizon remains regular with a finite and non-zero area, and its temperature becomes zero. It is interesting to note that in this limit, its entropy satisfies the very simple formula
\be
S=2\pi J_\phi\,.
\ee
Such a formula was first observed for the extremal Pomeransky--Sen'kov black ring by Reall \cite{Reall:2007jv}, and it led him to derive the entropy of that solution from a microscopic counting of states. A similar microscopic counting of states might apply to the present solution.

The dipole charge remains finite in this limit. It can be made to vanish if we set $\alpha=1/2$, and (\ref{extremal_limit}) reduces to the extremal Pomeransky--Sen'kov black ring, as expected. On the other hand, by fixing the ratio $0<\beta/\alpha<1$ and then taking the limit $\alpha\rightarrow 0$, we recover the extremal singly rotating dipole black ring studied in \cite{Emparan:2004wy} for Kaluza--Klein theory. However, we point out that in this singly rotating case, the black ring has a singular horizon with vanishing area.

Recall from the discussion of Sec.~5 that if we fix $j_{\psi}$ and $q$, the $S^2$-rotation $j_{\phi}$ is maximised in the extremal limit; on the other hand, if we fix $j_{\psi}$ and $j_{\phi}$ with $j_{\psi}+j_{\phi}>1$, the dipole charge $q$ is maximised in this limit. Hence the extremal black ring can be understood as a black-ring system which is either maximally $S^2$-rotating or maximally dipole-charged.

\newsubsection{Collapse limit}

In this limit, the ring radius shrinks down to zero, so the black ring ``collapses'' into a black hole. 
It is obtained by first fixing $a$ and $c$, setting
\be
b=\frac{1-a}{1+a}-\frac{\eta\varkappa^2}{1+a}\,,
\ee
and then taking $\varkappa\rightarrow 0$. In terms of the rod structure shown in Fig.~\ref{Final rod structure}, it corresponds to shrinking the lengths of both the second and third rods to zero, while keeping their ratio fixed. In this limit, we find a simple relation between the two angular momenta:
\be
j_{\psi}+j_{\phi}=1\,, \qquad \forall\, q\in\big[0,\hbox{$\sqrt{\frac{2}{3\pi}}$}\,\big].
\ee
If we ignore the existence of $q$, this relation describes the phase space of the five-dimensional extremal Myers--Perry black hole \cite{Emparan:2008eg}. In the phase space diagrams in Sec.~5, the collapse limit is represented by the dashed curves.

If we define the parameters $m$, $a_1$ and $a_2$ as
\be
m=\frac{4(a+c)}{\eta(1+a)}\,,\qquad a_1=\frac{\sqrt{2}(a^2c+2a+c)}{\sqrt{\eta(1+a)(a+c)}}\,,\qquad a_2=\frac{\sqrt{2(1+a)}c(1-a)}{\sqrt{\eta(a+c)}}\,,
\ee
and the coordinates $r$ and $\theta$ by
\be
x=-1+\frac{4\varkappa^2(1-c)\cos^2\theta}{r^2-a_1a_2}\,,\qquad y=-1-\frac{4\varkappa^2(1-c)\sin^2\theta}{r^2-a_1a_2}\,,
\ee
we find that the metric (\ref{solution5D}) reduces in this limit to the Myers--Perry solution:
\ba
\dif s^2&=&-\dif t^2+\frac{2m}{\Sigma}(\dif t-a_1\sin^2\theta\,\dif\psi-a_2\cos^2\theta\,\dif\phi)^2\cr
&&+(r^2+a_1^2)\sin^2\theta\,\dif\psi^2+(r^2+a_2^2)\cos^2\theta\,\dif\phi^2+\Sigma\left(\frac{\dif r^2}{\Delta}+\dif\theta^2\right),
\ea
where
\be
\Delta=r^2\bigg(1+\frac{a_1^2}{r^2}\bigg)\bigg(1+\frac{a_2^2}{r^2}\bigg)-2m\,,\qquad\Sigma=r^2+a_1^2\cos^2\theta+a_2^2\sin^2\theta\,.
\ee
At the same time, the dilaton (\ref{phi_soln}) decouples and the gauge potential (\ref{A_soln}) vanishes. 

In the zero dipole-charge case $a=c$, this limit was previously considered in \cite{Elvang:2007hs}. In the present case, although the parameter $a$ enters into the limiting metric non-trivially, there are essentially two independent parameters, as can be seen from the extremality condition $a_1+a_2=\sqrt{2m}$. This is of course expected: the extremal Myers--Perry black hole is not supposed to carry a third degree of freedom coming from dipole charge. However, it is interesting to note that not all the physical quantities are continuous in this limit, i.e., the physical quantities of our dipole ring in this limit may not be the same as those in the limiting extremal Myers--Perry solution. In particular, $q$ does {\it not\/} vanish in the collapse limit.

\newsubsection{Infinite ring-radius limit}
 
To take the infinite ring-radius limit of the doubly rotating dipole black ring, we first set 
\be
c=\frac{\sqrt{2(m^2-\alpha^2)}}{\varkappa}\,,\qquad
a=\frac{\sqrt{2(m^2-\alpha^2)}}{\varkappa}\,\cosh^2\gamma\,,\qquad
b=\frac{m-\sqrt{m^2-\alpha^2}}{m+\sqrt{m^2-\alpha^2}}\,,
\ee
and define the new coordinates $(r,\theta,z)$ by
\be
x=\cos\theta\,,\qquad
y=-\frac{\sqrt{2}\varkappa}{r-(m-\sqrt{m^2-\alpha^2})}\,,\qquad
\psi=\frac{z}{\sqrt{2}\varkappa}\,.
\ee
If we then take the $\varkappa\rightarrow\infty$ limit of the six-dimensional metric (\ref{solution6D}), we obtain:
\ba
\label{KK black string}
\dif s^2_6&=&\frac{\Sigma}{\Sigma+2mr\sinh^2\gamma}\bigg[\dif w'-\frac{2\alpha m\sinh\gamma\cos\theta}{\Sigma}\,\dif t'
+\frac{m\sinh2\gamma\cos\theta}{\Sigma-2mr}\cr
&&\times
\bigg(\Delta-\frac{2\alpha^2mr\sin^2\theta}{\Sigma}\bigg)\dif\phi\bigg]^2-\frac{\Sigma-2mr}{\Sigma}\left(\dif t'+\frac{2\alpha m\cosh\gamma\, r\sin^2\theta}{\Sigma-2mr}\,\dif\phi\right)^2\cr
&&+(\Sigma+2mr\sinh^2\gamma)\left(\frac{\dif r^2}{\Delta}+\dif\theta^2
+\frac{\Delta}{\Sigma-2mr}\sin^2\theta\,\dif\phi^2\right)
+\dif z'^2,
\ea
where
\be
\Sigma=r^2+\alpha^2\cos^2\theta\,,\qquad\Delta=r^2-2mr+\alpha^2.
\ee
Here, we have defined the coordinates:
\ba
\dif w'&=&\dif w+m\sinh2\gamma\,\dif\phi\,,\cr
\label{boost}
\dif t'&=&\cosh\sigma\,\dif t-\sinh\sigma\,\dif z\,,\cr
\dif z'&=&\cosh\sigma\,\dif z-\sinh\sigma\,\dif t\,,
\ea
where
\be
\label{boost_param_balance}
\sinh\sigma=\cosh\gamma\,.
\ee
When dimensionally reduced on $w'$ (or $w$), (\ref{KK black string}) describes a magnetically charged, rotating Kaluza--Klein black hole \cite{Rasheed:1995zv,Larsen:1999} that is extended along the $z'$-direction\footnote{If we write (\ref{KK black string}) in the form $\dif s^2_6=\dif s^2_5+\dif z'^2$, then (4.11) of \cite{Rasheed:1995zv} can be mapped to  $\dif s^2_5$ by setting $\alpha\mapsto0$, $\beta\mapsto\gamma$, $M_{\rm K}\mapsto m$, $a\mapsto\alpha$, $r\mapsto r-\Sigma/\sqrt{3}$, $t\rightarrow t'$ and $x^5\mapsto w'$.}---in other words, a Kaluza--Klein black string. Moreover, this solution is boosted along the $z'$-direction with a certain boost parameter $\sigma$, as can be seen from the last two equations of (\ref{boost}). In the limit when the rotation $\alpha$ goes to zero, we recover the charged non-rotating black string found in \cite{Emparan:2004wy}, for Kaluza--Klein theory and the boost parameter satisfying the balance condition (\ref{boost_param_balance}).

We remark that this result is entirely consistent with the interpretation made in Sec.~4 of the dipole black ring as a string of magnetically charged Kaluza--Klein black holes, bent into a circular shape.

\newsection{Discussion}

In this paper, we have used the ISM in six-dimensional vacuum gravity to construct a new solution of five-dimensional Kaluza--Klein theory describing a doubly rotating black ring carrying magnetic dipole charge. This black ring is balanced, so the space-time does not contain any conical singularities, and it can be regarded as a dipole-charged generalisation of the Pomeransky--Sen'kov black ring. We then studied some properties of this solution, including its phase space structure and its various limits.

We note that this solution is not the most general black ring possible in Kaluza--Klein theory. It is clearly possible to generalise it to the unbalanced case, with independent rotations in both directions. Indeed, it is possible to obtain this solution from the ISM construction of Sec.~2, by relaxing the conditions made in (\ref{z1andC}) so that Rods 1 and 4 (after dimensional reduction) are parallel only. However, based on our experience with the most general vacuum black ring \cite{Chen:2011jb}, this solution is likely to be very complicated. It might be worthwhile to focus on obtaining another special case of this solution in the first instance, namely the purely $S^2$-rotating dipole ring. This would be the dipole-charged generalisation of the solution in \cite{Mishima:2005id,Figueras:2005zp}.

Even if we restrict ourselves to the balanced case, the solution in this paper is still not the most general one possible in Kaluza--Klein theory. Such a distinction would belong to a doubly rotating black ring that carries {\it both\/} magnetic dipole charge and a normal conserved electric charge. It may be possible to add a conserved electric charge to the present solution by applying one of the standard charging transformations, although the resulting space-time is likely to contain Dirac--Misner singularities. A better approach, following the original spirit of how the present solution was derived, would be to use the ISM on a suitable seed to generate this most general balanced solution. We leave this interesting problem for the future.

A more ambitious task would be to find doubly rotating dipole black ring solutions to the action (\ref{5D_action}), for other values of dilaton coupling $\alpha$. The method developed in \cite{Rocha:2011vv} and used in this paper will no longer be applicable, although it is conceivable that some variation of the ISM might be applicable for other special values of $\alpha$.
One particularly interesting case is pure Einstein--Maxwell theory with $\alpha=0$. It is also possible to move beyond Einstein--Maxwell theory by adding a Chern--Simons term, to obtain the bosonic sector of five-dimensional minimal supergravity theory. An inverse-scattering formalism \cite{Figueras:2009mc} was recently developed for this theory, and it perhaps offers the best hope of generating another example of a doubly rotating dipole black ring. Such a solution, if found, could be the starting point to generate the most general stationary black-ring solution of $U(1)^3$ supergravity theory \cite{Elvang:2004xi}.

Returning to five-dimensional Kaluza--Klein theory, it is notable that this theory can be interpreted as the NS-NS sector of low-energy string theory after the dualisation (\ref{duality}). Thus, our solution, and its generalisations discussed above, admit an embedding in string theory with rather interesting implications. In particular, when the extremal limit of our solution is taken, it describes a loop of fundamental string, rotating in both the loop direction as well as the direction orthogonal to it. It might be worthwhile to study this system and its qualitative microscopics in more detail.

\bigbreak\bigskip\bigskip\centerline{{\bf Acknowledgements}}
\nobreak\noindent ET wishes to thank Susan Scott and members of the Centre for Gravitational Physics, where this work was completed, for their kind hospitality. This work was partially supported by the Academic Research Fund (WBS No.: R-144-000-277-112) from the National University of Singapore.

\bigskip\bigskip

{\renewcommand{\Large}{\normalsize}
}

\end{document}